\documentclass[12pt]{article}

\usepackage{amssymb}
\usepackage{amsmath}
\usepackage{amscd}
\usepackage{latexsym}
\usepackage{graphicx}
\usepackage{caption}

\usepackage{cite}

\usepackage{amssymb}
\usepackage{amsmath}
\usepackage{amscd}
\usepackage{latexsym}
\usepackage{graphicx}

\newcommand{\be}{\begin{equation}}
\newcommand{\ee}{\end{equation}}

\newcommand{\dlt}{\delta}
\newcommand{\prt}{\partial}
\newcommand{\br}{{\bf r}}
\newcommand{\bB}{{\bf B}}
\newcommand{\bS}{{\bf S}}
\newcommand{\cH}{{\cal H}}
\newcommand{\cA}{{\cal A}}
\newcommand{\cD}{{\cal D}}
\newcommand{\bt}{\beta}
\newcommand{\vp}{\varphi}

\newcommand{\al}{\alpha}
\newcommand{\ra}{\rightarrow}

\newcommand{\gm}{\gamma}

\newcommand{\Om}{\Omega}

\newcommand{\dgr}{\dagger}

\newcommand{\rgl}{\rangle}
\newcommand{\lgl}{\langle}

\begin{document}

\begin{center}

{\Large{\bf Zeroth-Order Nucleation Transition under Nanoscale Phase 
Separation} \\ [5mm]

V.I. Yukalov$^{1,2}$ and E.P. Yukalova$^{3}$ }  \\ [3mm]

{\it
$^1$Bogolubov Laboratory of Theoretical Physics, \\
Joint Institute for Nuclear Research, Dubna 141980, Russia \\ [2mm]

$^2$Instituto de Fisica de S\~ao Carlos, Universidade de S\~ao Paulo, \\
CP 369, S\~ao Carlos 13560-970, S\~ao Paulo, Brazil \\ [2mm]

$^3$Laboratory of Information Technologies, \\
Joint Institute for Nuclear Research, Dubna 141980, Russia } \\ [3mm]

{\bf E-mails}: {\it yukalov@theor.jinr.ru}, ~~ {\it yukalova@theor.jinr.ru}

\end{center}

\vskip 1cm

\begin{abstract}
Materials with nanoscale phase separation are considered. A system 
representing a heterophase mixture of ferromagnetic and paramagnetic phases is 
studied. After averaging over phase configurations, a renormalized Hamiltonian 
is derived describing the coexisting phases. The system is characterized by 
direct and exchange interactions and an external magnetic field. The properties 
of the system are studied numerically. The stability conditions define the stable
state of the system. At a  temperature of zero, the system is in a pure ferromagnetic 
state. However, at finite temperature, for some interaction parameters, the system 
can exhibit a zeroth-order nucleation transition between the pure ferromagnetic 
phase and the mixed state with coexisting ferromagnetic and paramagnetic phases.
At the nucleation transition, the finite concentration of the paramagnetic phase 
appears via a jump. 
\end{abstract}

\vskip 2cm
{\parindent=0pt
{\bf Keywords}: nanoscale phase separation; quasi-equilibrium system; 
heterophase mixture; zeroth-order transition; nucleation point  }

\newpage

\section{Introduction}

Phase transitions are commonly characterized by the appearance of 
non-analyticities in the system's thermodynamic characteristics. The 
classification of phase transitions is usually connected with the 
non-analyticities in the derivatives of thermodynamic potentials. Thus, 
the non-analyticity in the first-order derivatives implies a first-order 
phase transition, the non-analyticity in the second-order derivatives of 
a thermodynamic potential defines a second-order phase transition 
\cite{Kubo_1}.  

Recently, the possible existence of zeroth-order phase transitions has been 
brought to attention \cite{Maslov_2}; it is possible when a thermodynamic potential itself 
exhibits a discontinuity. Zeroth-order phase transitions have been found in 
the physics of black holes \cite{Gunasekaran_3,Altamirano_4,Altamirano_5,
Frassino_6,Hennigar_7,Hennigar_8,Kubiznak_9,Dehyadegari_10,Dehyadegari_11}, 
holographic superconductors \cite{Zeng_12,Cai_13,Cai_14,Zeng_15}, and 
holographic ferromagnets and antiferromagnets \cite{Cai_16,Cai_17}. The 
zeroth-order phase transition was also found for some spin models with 
long-range interactions \cite{Hou_18,Hou_19}. Note that for the systems 
with long-range interactions, microcanonical and canonical ensembles are 
not necessarily equivalent \cite{Touchette_20}. 

Metal-insulator phase transitions in some materials, such as V$_2$O$_3$,
were classified as zeroth-order phase transitions, where the free energy 
is discontinuous \cite{Bar_21,Kundu_22}. These phase transitions exhibit 
the phase coexistence and ramified fractal-like nanoscale phase separation 
in the transition region \cite{Bar_21,Kundu_22,Liu_23}.  
  
In this way, the zeroth-order phase transitions can occur when at least 
one of the features is present: either long-range interactions or nanoscale 
phase separation. Under this kind of phase separation, the system represents 
a mixture of nanoscale regions of different phases. The probabilistic weights 
of the phases are self-consistently defined by the system parameters and 
thermodynamic variables. Such nanoscale mixtures are also called heterophase 
or mesoscopic, since the linear size of inclusions of one phase inside the 
matrix of the other is larger than the interparticle distance but much smaller 
than the system linear size. The appearance of mesoscopic heterophase mixtures 
under nanoscale phase separation is a very widespread phenomenon arising around 
many phase transitions that can be of first or second order. Numerous examples 
of materials exhibiting the existence of such mixtures are given in the review 
articles \cite{Yukalov_24,Yukalov_25,Yukalov_26,Kagan_27}. Recently, the 
possibility of superfluid dislocations inside quantum crystals has been 
discussed \cite{Kuklov_43,Boninsegni_44,Yukalov_45}. Different types of nanoscale 
phase separation occur in electrolytes \cite{Buggy_1,Batys_2,Park_3,Martin_4,Kyndiah_5}. 

Here we shall concentrate on a heterophase mixture of ferromagnetic and
paramagnetic phases. There exist numerous examples of materials exhibiting the 
coexistence of magnetic (ferromagnetic or antiferromagnetic) and paramagnetic phases. 
Thus, using the M\"{o}ssbauer effect, the coexistence of antiferromagnetic and 
paramagnetic phases is observed in FeF$_3$ \cite{Bertelsen_30}, in CaFe$_2$O$_4$ 
\cite{Yamamoto_31}, and in a number of orthoferrites, such as LaFeO$_3$, PrFeO$_3$, 
NdFeO$_3$, SmFeO$_3$, EuFeO$_3$, GdFeO$_3$, TbFeO$_3$, DyFeO$_3$, YFeO$_3$,
HoFeO$_3$, ErFeO$_3$, TmFeO$_3$, and YbFeO$_3$ \cite{Eibschutz_32,Levinson_33}. 
Ferromagnetic cluster fluctuations, called ferrons or fluctuons, can arise inside 
a paramagnetic matrix of some semiconductors 
\cite{Krivoglaz_34,Nagaev_35,Belov_36,Belov_37}. In some materials, 
magnetic cluster excitations can occur in the paramagnetic region above $T_c$ or 
above $T_N$ \cite{Goldman_39,Reimann_40,Bhargava_41,Uen_42,Srivastava_43,Halg_44}, 
causing the appearance of spin waves in the paramagnetic phase, for instance, in Ni, 
Fe, EuO, EuS, Pd$_3$Fe, and Gd \cite{Lynn_45,Liu_46,Lynn_47,Cable_48,Lynn_49,Cable_50}. 
The coexistence of ferromagnetic and nonmagnetic phases was also observed in Y$_2$Co$_7$, 
YCo$_3$, Co(S$_x$Se$_{1-x}$)$_2$, Co(Ti$_x$Al$_{1-x}$)$_2$, and Lu(Co$_{1-x}$Al$_x$)$_2$ 
\cite{Goto_51,Shinogi_52}. In colossal magnetoresistance materials, such as 
La$_{1-x}$Ca$_x$MnO$_3$ and La$_{1-x}$Sr$_x$CoO$_3$, one observes the coexistence of a
paramagnetic insulating, or semiconducting, phase and a ferromagnetic metallic phase
\cite{Jaime_53,Merithew_54,Baio_55}, while in La$_{0.67-x}$Bi$_x$Ca$_{0.33}$MnO$_3$, 
paramagnetic and antiferromagnetic phases coexist \cite{Sun_56}. Nanoscale phase 
separation into ferromagnetic and paramagnetic regions has been observed in the colossal 
magnetoresistence compound, EuB$_{5.99}$C$_{0.01}$ \cite{Batko_68}. Many more examples
can be found in the review articles \cite{Yukalov_24,Yukalov_25,Yukalov_26,Kagan_27}.
 %MDPI: refs. 44--46 are missing, please check
 
In the present paper, we consider a heterophase system with random phase 
separation, where the regions of different phases are randomly distributed 
in space. By averaging the phase configurations, we derive a renormalized, 
effective Hamiltonian of the mixture. Keeping in mind a spin system, we pass 
to the quasi-spin representation. Specifically, we consider a mixture of 
ferromagnetic and paramagnetic phases. Long-range interactions are assumed, 
such that the mean-field approximation becomes, in the thermodynamic limit, 
asymptotically exact. The existence of the ferromagnetic--paramagnetic mixture 
is due to the competition between direct and exchange interactions. We treat 
the case when the system is placed in an external magnetic field. We show 
that for some system parameters, there occurs the following situation: at low 
temperatures, the system is a pure ferromagnet that, when rising in temperature, 
can transfer into a mixture of ferromagnetic and paramagnetic phases at a 
nucleation point. For some system parameters, this nucleation transition 
happens to be a zeroth-order transition. 

The plan of the paper is as follows. In Section \ref{sect:two}, we recall the Gibbs method 
of equimolecular surfaces that are used for describing the spatial phase 
separation. Section \ref{sect:three} explains how the statistical operator of the mixture 
with phase separation can be defined by minimizing the functional information.%please check intended meaning is retained.
The random spatial distribution of competing phases requires the averaging 
over phase configurations. The results of this averaging are summarized in 
Section \ref{sect:four}. In Section \ref{sect:five}, we pass from the field-operator representation to spin 
representation. Although this conversion is based on the known Bogolubov 
canonical transformation, it is necessary to recall it in order to elucidate 
the importance of taking account of direct particle interactions, in addition 
to exchange interactions. Keeping in mind long-range interactions, in Section 
\ref{sect:six}, we derive the free energy of the mixture. \mbox{Section \ref{sect:seven}} 
formulates the stability conditions that make it straightforward to separate stable states 
from unstable ones. In Section \ref{sect:eight}, we present the results of the numerical 
calculations and accompany them with discussions and conclusions.

\section{Spatial Phase Separation}\label{sect:two}

The description of a two-phase system with spatial phase separation starts 
with the Gibbs method \cite{Gibbs_28} of equimolecular separating surfaces, 
where the system of volume $V$ and number of particles $N$ is considered to 
be separated into two parts, with the total volumes $V_f$ and the particle 
numbers $N_f$, so that
\be
\label{1}
 V = V_1 + V_2 \; , \qquad N = N_1 + N_2 \; .
\ee

The regions $\mathbb{V}_f$ occupied by different phases are assumed to be 
randomly distributed in space. Their spatial locations are described by the 
manifold indicator functions
\begin{eqnarray}
\label{2}
\xi_f(\br) = \left\{ \begin{array}{ll}
1 \; , ~ & ~ \br \in \mathbb{V}_f \\
0 \; , ~ & ~ \br \not\in \mathbb{V}_f 
\end{array}  \right. \; ,
\end{eqnarray}
where
$$
V_f \equiv {\rm mes} \; \mathbb{V}_f \qquad ( f = 1,2 ) \; .
$$

The Hilbert space of microscopic states of the system is the tensor product
\be
\label{3}
\cH = \cH_1 \bigotimes \cH_2
\ee
of the weighted Hilbert spaces \cite{Yukalov_24,Yukalov_25,Yukalov_26} 
corresponding to the phases $f=1,2$. The algebra of observables in this 
space is given by the direct sum of the algebra representations on the 
corresponding subspaces
\be
\label{4}
\cA(\xi) =  \cA_1 (\xi_1)  \bigoplus \cA_2 (\xi_2) \; .
\ee

For instance, the system energy Hamiltonian reads as
\be
\label{5}
 \hat H(\xi) =  \hat H_1 (\xi_1)  \bigoplus \hat H_2 (\xi_2) \;  ,
\ee
with the general form of the phase replica Hamiltonians
$$
\hat H_f(\xi_f) = \int \xi_f(\br) \psi_f^\dgr(\br) 
\left[ \; -\; \frac{\nabla^2}{2m} + U(\br) \; 
\right] \; \psi_f(\br) \; d\br  +
$$
\be
\label{6}
+
\frac{1}{2} \int \xi_f(\br)\; \xi_f(\br') \; 
\psi_f^\dgr(\br)\; \psi_f^\dgr(\br') \; \Phi(\br-\br')\; 
\psi_f(\br') \; \psi_f(\br)\; d\br d\br' \;  ,
\ee
where $\Phi({\bf r})$ is an interaction potential, $U({\bf r})$ is an external 
potential, and the field operators $\psi_f({\bf r})$ are columns with respect 
to internal degrees of freedom, such as spin.  The number-of-particle operator 
is
\be
\label{7}
 \hat N(\xi) =  \hat N_1 (\xi_1) +  \hat N_2 (\xi_2) \;  ,
\ee
with the number-of-particle operators of each phase
\be
\label{8}
 \hat N_f(\xi_f) = 
\int \xi(\br)\; \psi_f^\dgr(\br)\; \psi_f(\br)\; d\br \; .
\ee

Here and below, we set  the Planck and Boltzmann constants to one.

\section{System Statistical Operator}\label{sect:three}

The general procedure of defining the statistical operator for a system is 
by minimizing the information functional, taking account of the prescribed 
constraints. The latter is the normalization condition
\be
\label{9}
{\rm Tr} \; \int \hat\rho(\xi) \; \cD\xi = 1 \;  ,
\ee
the definition of the system energy
\be
\label{10}
{\rm Tr} \; \int \hat\rho(\xi)\; \hat H(\xi) \; \cD\xi = E \; ,
\ee
and of the total number of particles in the system
\be
\label{11}
{\rm Tr} \; \int \hat\rho(\xi)\; \hat N(\xi) \; \cD\xi = N \;   .
\ee

Here and in what follows, the trace operation is taken over the whole 
Hilbert space (\ref{3}), and $\mathcal{D}\xi$ implies the averaging over 
phase configurations describing the random locations and shapes of 
separated phases. 

The information functional in the Kullback--Leibler form 
\cite{Kullback_29,Kullback_30} reads as
$$
I[\; \hat\rho\; ] = {\rm Tr} \; \int \hat\rho(\xi)\; 
\ln \; \frac{\hat\rho(\xi)}{\hat\rho_0(\xi)} \; \cD\xi \; + \; 
\al \left[ \; {\rm Tr} \; \int \hat\rho(\xi)\; \cD\xi - 1 \; 
\right] \; +
$$
\be
\label{12}
 + \; 
\bt  \left[ \; {\rm Tr} \; 
\int \hat\rho(\xi)\; \hat H(\xi) \; \cD\xi - E \; 
\right] \; + \;  
\gm  \left[ \; {\rm Tr} \; 
\int \hat\rho(\xi)\; \hat N(\xi) \; \cD\xi - N \; 
\right] \;         ,
\ee
with the Lagrange multipliers $\al$, $\bt=1/T$, and $\gm=-\bt\mu$, and 
with a trial statistical operator $\hat{\rho}_0(\xi)$ characterizing 
some a priori information if any. If no a priori information is available, 
$\hat{\rho}_0(\xi)$ is a constant. Then minimizing the information functional 
over $\hat{\rho}(\xi)$ yields the statistical operator
\be
\label{13}
\hat\rho(\xi) = \frac{1}{Z} \; \exp\{ - \bt H(\xi) \} \; ,
\ee
with the grand Hamiltonian 
\be
\label{14}  
 H(\xi) = \hat  H(\xi) - \mu  \hat N(\xi)
\ee
and the partition function
\be
\label{15}
Z = {\rm Tr} \; \int \exp\{ - \bt H(\xi) \} \; \cD\xi \;  .
\ee

Introducing the effective renormalized Hamiltonian by the relation
\be
\label{16}
\exp\{ - \bt \widetilde H \} = \int \exp\{ - \bt H(\xi) \} \; \cD\xi
\ee 
gives the partition function
\be
\label{17}
Z = {\rm Tr} \; \exp\{ - \bt \widetilde H \} \; .
\ee

Then we get the grand thermodynamic potential
\be
\label{18}
  \Om = - T \ln Z \;  .
\ee

This picture describes a heterophase system where the phase-separated regions 
are random in the sense that they are randomly located in space and can move and 
change their shapes. In that sense, strictly speaking, the system is in 
quasi-equilibrium. However, the averaging over phase configurations reduces 
the consideration to an effective system equilibrium on average 
\cite{Yukalov_24,Yukalov_25,Yukalov_26}.

\section{Averaging over Phase Configurations}\label{sect:four}

In order to explicitly accomplish  the averaging over phase configurations, 
it is necessary to define the functional integration over the manifold 
indicator functions (\ref{2}). This functional integration has been defined 
and explicitly realized in papers 
\cite{Yukalov_24,Yukalov_31,Yukalov_32,Yukalov_33,Yukalov_34,Yukalov_35}. 
Here we formulate the main results of this functional integration over the 
manifold indicator functions with the differential measure $\cD\xi$,   
which realizes the averaging over phase configurations. 
 
\vskip 3mm
{\bf Theorem 1.}
Let us consider the functional
\be
\label{19}
A_f(\xi_f) = \sum_{n=0}^\infty \int \xi_f(\br_1)\; \xi_f(\br_2)\ldots
\xi_f(\br_n ) \; A_f(\br_1,\br_2,\ldots,\br_n) \; 
d\br_1 d\br_2 \ldots d\br _n \; .
\ee

The integration of this function over the manifold indicator functions 
gives
\be
\label{20}
 \int A_f(\xi_f) \;\cD\xi = A_f(w_f) \; ,
\ee
where
\be
\label{21}
 A_f(w_f) = \sum_{n=0}^\infty w_f^n 
\int A_f(\br_1,\br_2,\ldots,\br_n) \; d\br_1 d\br_2 \ldots d\br _n \; ,   
\ee
while 
\be
\label{22}
w_f = \frac{1}{V} \int \xi_f(\br) \; d\br = \frac{V_f}{V}
\ee
defines the geometric probability of an $f$-th phase. 

\vskip 2mm
{\bf Theorem 2.}
The thermodynamic potential
\be
\label{23}
 \Om = - T \ln \; {\rm Tr}\; \int \exp\{ - \bt H(\xi) \} \; \cD\xi \; ,
\ee
after the averaging over phase configurations, becomes
\be
\label{24}
\Om = - T \ln \; {\rm Tr}\; \{ -\bt \widetilde H \} = \sum_f \Om_f \equiv 
\Om(w) \; ,
\ee
where
\be
\label{25}
 \Om_f = - T \ln \; {\rm Tr}_{\cH_f}\; \{ -\bt H_f(w_f) \} \equiv 
\Om_f(w_f) \; ,
\ee
and the renormalized Hamiltonian is
\be
\label{26}
 \widetilde H = \bigoplus_f H_f(w_f) \equiv \widetilde H(w) \; ,
\ee
with the phase probabilities $w_f$ being the minimizers of the thermodynamic 
potential,
\be
\label{27}
 \Om = {\rm abs} \;\min_{\{ w_f \} } \; \Om(w) \; ,
\ee
under the normalization condition
\be
\label{28}
 \sum_f w_f = 1 \; , \qquad 0 \leq w_f \leq  1 \; .
\ee

{\bf Theorem 3.}
The observable quantities, given by the averages
\be
\label{29}
 \lgl \; \hat A \; \rgl =
{\rm Tr} \; \int \hat\rho(\xi) \;  \hat A(\xi) \; \cD\xi 
\ee
of the operators from the algebra of observables (\ref{4}),
\be
\label{30}
\hat A(\xi) = \bigoplus_f  \hat A_f(\xi_f) \; ,
\ee
with $\hat{A}_f(\xi_f)$ defined as in (\ref{19}), after the averaging over 
phase configurations, reduce to \mbox{the form}
\be
\label{31}
\lgl \; \hat A \; \rgl = {\rm Tr} \; \hat\rho(w) \; \hat A(w) \; ,
\ee
where the renormalized operator of an observable is
\be
\label{32}
 \hat A(w) = \bigoplus_f \hat A_f(w_f) \; ,
\ee
with $\hat{A}_f(w_f)$ defined as in (\ref{21}), and the renormalized 
statistical operator is
\be
\label{33}
 \hat\rho(w) = \frac{1}{Z} \; \exp\{ - \bt \widetilde H(w) \}  \; ,
\ee
with the partition function (\ref{17}).

The proofs of these theorems are given in the papers 
\cite{Yukalov_24,Yukalov_31,Yukalov_32,Yukalov_33,Yukalov_34,Yukalov_35}.

\section{Hamiltonian in Spin Representation}\label{sect:five}

Since we aim to study the magnetic properties of a system with phase 
separation, it is useful to transform Hamiltonian (\ref{6}) into spin 
representation. For this purpose, we assume that the system is periodic 
over a lattice with the lattice sites $\br_j$, where $j=1,2,\ldots,N$, 
and we expand the field operators over Wannier functions:
\be
\label{34}
\psi_f(\br) = \sum_j c_{jf}\vp_f(\br - \br_j) \;  .
\ee

Keeping in mind well-localized Wannier functions \cite{Marzari_36}, we 
retain in the Hamiltonian only the terms expressed through the matrix 
elements over Wannier functions containing not more than two lattice sites, 
since the overlap of Wannier functions located at three or four different 
lattice sites is negligibly small. 

The remaining matrix elements are: the tunneling term
\be
\label{35}
T_{ijf} = - \int \vp_f^*(\br-\br_i) \; \left[\; - \; \frac{\nabla^2}{2m} +
U(\br) \; \right] \; \vp_f(\br-\br_j) \;  d\br \; ,
\ee
the term of direct interactions
\be
\label{36}
\Phi_{ijf} = \int |\; \vp_f(\br-\br_i) \; |^2 \; \Phi(\br-\br') \;
 |\; \vp_f(\br'-\br_j) \; |^2 \; d\br d\br' \;  ,
\ee
and the term of exchange interactions
\be
\label{37}
J_{ijf} =  - \int \vp_f^*(\br-\br_i) \; \vp_f^*(\br'-\br_j) \; 
\Phi(\br-\br') \;
 \vp_f(\br'-\br_i) \; \vp_f(\br-\br_j) \; d\br d\br' \; .
\ee

Then the Hamiltonian (\ref{6}) transforms into the form
$$
H_f = 
- w_f \sum_{ij} ( T_{ijf} + \mu\dlt_{ij} ) c_{if}^\dgr \; c_{jf} \; +
$$
\be
\label{38}
+ \; \frac{1}{2} \; w_f^2
\sum_{ij} \left( \Phi_{ijf} \; c_{if}^\dgr\; c^\dgr_{jf} \;
 c_{jf}\; c_{if} - J_{ijf} \; c_{if}^\dgr\; c^\dgr_{jf} \;
 c_{if}\; c_{jf} \right) \; .
\ee

To exclude self-interactions, one sets
\be
\label{39}
 \Phi_{jjf}  \equiv J_{jjf} = 0 \;.
\ee

Then we introduce spin operators following the method of canonical \linebreak
transformations \cite{Bogolubov_37,Bogolubov_38,Bogolubov_39}, generalized 
in the case of heterophase systems \cite{Yukalov_40,Yukalov_41,Yukalov_42}. 
Keeping in mind the particles with spin one-half, the operators $c_{jf}$ are 
to be treated as spinors
\begin{eqnarray}
\label{40}
c_{jf} = \left[ \begin{array}{c}
c_{jf}(\uparrow) \\
\\
c_{jf}(\downarrow) \end{array} \right]
\end{eqnarray}
of two components, one with spin up and the other with spin down. When each 
lattice site is occupied by a single particle, the unipolarity condition is 
valid
\be
\label{41}
c_{jf}^\dgr(\uparrow) \; c_{jf}(\uparrow) + 
c_{jf}^\dgr(\downarrow) \; c_{jf}(\downarrow) = 1 \;  .
\ee

The canonical transformations introducing spin operators ${\bf S}_{jf}$, 
acting on the space $\mathcal{H}_f$, read as
$$
c_{jf}^\dgr(\uparrow) \; c_{jf}(\uparrow) = 
\frac{1}{2} + S_{jf}^z \; ,
\qquad  
c_{jf}^\dgr(\uparrow) \; c_{jf}(\downarrow) = 
S_{jf}^x + i \; S_{jf}^y \;
$$
\be
\label{42}
c_{jf}^\dgr(\downarrow) \; c_{jf}(\downarrow) = 
\frac{1}{2}\; - \; S_{jf}^z \;  .
\ee

Employing these canonical transformations and wishing to write the 
Hamiltonian in a compact form, we define the average direct interactions
\be
\label{43}
\Phi_f \equiv \frac{1}{N} \sum_{i\neq j} \Phi_{ijf}  \;  ,
\ee
the average exchange interactions
\be
\label{44}
J_f \equiv \frac{1}{N} \sum_{i\neq j} J_{ijf}  \; ,
\ee
and the effective chemical potentials
\be
\label{45}
\mu_f \equiv \mu +  \frac{1}{N} \sum_{ij} T_{ijf}  \;  .
\ee

Then Hamiltonian (\ref{38}) becomes
\be
\label{46}
 H_f = \frac{1}{2} \; w_f^2 U_f N - 
w_f^2 \sum_{i\neq j} J_{ijf} \bS_{if} \cdot \bS_{jf} - w_f \mu_f N \; ,
\ee
where
\be
\label{47}
 U_f \equiv \Phi_f - \; \frac{1}{2}\; J_f \;  .
\ee

For localized particles, the tunneling term $T_{ijf}$ is small and can be 
neglected. Hence, as is seen from expression (\ref{45}), $\mu_f=\mu$. Then 
the last term in Hamiltonian (\ref{46}) becomes $-w_f \mu N$. Such linear scalar 
terms  in $w_f$ can be omitted since they enter the\linebreak Hamiltonian 
(\ref{26}) as $w_1\mu+w_2\mu=\mu$, which is as a constant shift. The value 
(\ref{47}) characterizes an average potential acting on each particle in 
the system and is mainly due to direct interactions that are usually much 
larger than the exchange interactions. It is reasonable to assume that this 
average potential does not depend on the kind of magnetic phases, so that 
$U_f = U$. For generality, it is also necessary to take into account an 
external magnetic field ${\bf B}_0$. As a result, we come to the Hamiltonian
\be
\label{48}
H_f = \frac{1}{2} \; w_f^2 U N - 
w_f^2 \sum_{i\neq j} J_{ijf} \bS_{if} \cdot \bS_{jf} -
w_f \sum_j \mu_0 \bB_0 \cdot \bS_{jf} \;  .
\ee
 
The main feature of the paramagnetic phase is, clearly, the absence of long-range order. 
The direct way of taking this into account on the microscopic level is to notice that the 
term of exchange interactions (\ref{37}) essentially depends on the localization of 
Wanier functions. From expression (\ref{37}), it is evident that the better Wannier 
functions are localized, the smaller the exchange term. Therefore, accepting that the 
paramagnetic exchange term is very small, automatically degrades the long-range order.
Keeping this in mind, we set to zero the paramagnetic exchange interactions, $J_{ij2}=0$. 
Then the Hamiltonian (\ref{48}) yields for the ferromagnetic phase
\be
\label{49}
H_1 = \frac{1}{2} \; w_1^2 U N - 
w_1^2 \sum_{i\neq j} J_{ij1} \bS_{i1} \cdot \bS_{j1} -
w_1 \sum_j \mu_0 \bB_0 \cdot \bS_{j1} \; ,
\ee
and for the paramagnetic phase
\be
\label{50}
H_2 = 
\frac{1}{2} \; w_2^2 U N - w_2 \sum_j \mu_0 \bB_0 \cdot \bS_{j2} \; .
\ee

The external magnetic field is assumed to be directed along the $z$-axis
\be
\label{51}
 \bB_0 = B_0 {\bf e}_z \qquad ( B_0 \geq 0 ) \; .
\ee

Recall that the total system Hamiltonian, according to (\ref{26}), reads as
\be
\label{52}
 \widetilde H = H_1 \bigoplus H_2 \; .
\ee

The order parameters can be defined by the averages
\be
\label{53}
s_f \equiv \left\lgl \; \frac{2}{N} \sum_j S_{jf}^z \; \right\rgl \; ,
\ee
which lie in the interval
\be
\label{54}
 0 \leq s_f \leq 1 \qquad ( f = 1,2 ) \; .
\ee

For the ferromagnetic phase, there exist such low temperatures where
\be
\label{55}
\lim_{B_0\ra 0} s_1 > 0 \qquad ( T \ra 0 ) \; ,
\ee
while for the paramagnetic phase at all temperatures, one has
\be
\label{56}
\lim_{B_0\ra 0} s_2 = 0 \;  .
\ee

Accepting that interparticle interactions are of a long-range order, 
Hamiltonian (\ref{49}) can be simplified by resorting to the mean-field 
approximation
\be
\label{57}
 S_{i1}^z S_{j1}^z = S_{i1}^z \; \lgl \; S_{j1}^z \; \rgl + 
\lgl \; S_{i1}^z \; \rgl \; S_{j1}^z - 
\lgl \; S_{i1}^z \; \rgl \lgl \; S_{j1}^z \; \rgl \; ,
\ee
where $i \neq j$. This reduces that Hamiltonian to the form
\be
\label{58}
 H_1 = \frac{1}{2}\; w_1^2 \left( 
U + \frac{1}{2} \; J s_1^2 \right) N -
\left( w_1^2 J s_1 + w_1 \mu_0 B_0 \right) \sum_j S_{j1}^z \;  .
\ee

\section{Free Energy of Mixture} \label{sect:six}  

Defining the reduced free-energy in the standard way
\be
\label{59}
F = - \; \frac{ T}{N} \; \ln \; {\rm Tr} \; e^{-\bt\widetilde H} \; ,
\ee
introducing the dimensionless parameters
\be
\label{60}
u \equiv \frac{U}{J} \; , \qquad h \equiv \frac{\mu_0B_0}{J} \; ,
\ee
where
\be
\label{61}
 J \equiv J_1 = \frac{1}{N} \sum_{i\neq j} J_{ij1} \; ,
\ee
and measuring temperature in units of $J$, we come to the mixture free 
energy
\be
\label{62}
F = F_1 + F_2 \; .
\ee

Here the free energy of the magnetic component is
\be
\label{63}
F_1 = \frac{1}{2} \; w_1^2 \; \left( u + \frac{1}{2} \; s_1^2\right) -
T \; \ln \left[\; 
2 \cosh\left( \frac{w_1h+w_1^2 s_1}{2T} \right)\; \right] \; ,
\ee
with the order parameter
\be
\label{64}
 s_1 = \tanh\left( \frac{w_1 h + w_1^2 s_1}{2T} \right) \; ,
\ee
and the free energy of the paramagnetic component is
\be
\label{65}
 F_2 = \frac{1}{2} \; w_2^2 \; u - T \; \ln \left[\; 
2 \cosh\left( \frac{w_2h}{2T} \right)\; \right] \; ,
\ee
with the order parameter
\be
\label{66}
 s_2 = \tanh\left( \frac{w_2 h }{2T} \right) \; .
\ee

Studying the properties of the free energy, it is convenient to represent 
it in the form symmetric with respect to both phase components, introducing 
the quantity
\be
\label{67}
 g_f \equiv \frac{1}{N} \; \sum_{i\neq j} \frac{J_{ijf}}{J} \; .
\ee

By definition, $g_1=1$, while $g_2\ra 0$. Then the partial free energy 
\be
\label{68}
F_f = -\; \frac{T}{N} \; \ln \; {\rm Tr} \; e^{-\bt H_f}   
\ee
becomes
\be
\label{69}
F_f = \frac{1}{2} \; w_f^2 \; \left( u + \frac{1}{2} \; g_f s_f^2\right) -
T \; \ln \left[\; 
2 \cosh\left( \frac{w_fh+w_f^2 g_f s_f}{2T} \right)\; \right] \;   ,
\ee
with the order parameter
\be
\label{70}
 s_f = \tanh\left( \frac{w_f h + w_f^2 g_f s_f}{2T} \right) \;  .
\ee

\section{Stability Conditions}\label{sect:seven}

The statistical system is stable when it is in the state of the absolute 
minimum of the thermodynamic potential, which in the present case is the free 
energy. The system is in the mixed state, provided the free energy (\ref{62}) 
corresponds to a minimum with respect to the variables $w_1$, $s_1$, and $s_2$. 
The variable $w_2$ is expressed through the relation $w_2 = 1- w_1$. For 
convenience, it is possible to use the notation
\be
\label{71}
 w_1 \equiv w \; , \qquad w_2 = 1 - w  
\ee
and consider only the variable $w$, instead of $w_1$ and $w_2$ connected by 
the normalization condition. The conditions of the extremum are
\be
\label{72}
\frac{\prt F}{\prt s_1} = 0 \; , \qquad 
\frac{\prt F}{\prt s_2} = 0 \; , \qquad 
\frac{\prt F}{\prt w} = 0 \;  .
\ee

The first and second conditions give the expressions (\ref{64}) and (\ref{66}) 
for the order parameters $s_1$ and $s_2$. The third equation, due to the 
normalization condition, can be \mbox{written as}
$$
\frac{\prt F}{\prt w_1} = \frac{\prt F}{\prt w_2} \;  .
$$

Using the derivative
$$
\frac{\prt F}{\prt w_f} = w_f \left( u - \; 
\frac{1}{2} \; g_f s_f^2 \right) - \; \frac{1}{2} \; h s_f
$$
results in the probability of the ferromagnetic component
\be
\label{73}
 w = \frac{2u+h(s_1-s_2)}{4u-s_1^2} \; .
\ee
   
The extremum is a minimum provided the principal minors of the Hessian 
matrix are positive. The Hessian matrix is expressed through the second 
derivatives
$$
\frac{\prt^2 F_f}{\prt w_f^2} = u - \; \frac{1}{2}\; g_f s_f^2 - \;
\frac{1-s_f^2}{4T} \; \left( h + 2w_f g_f s_f \right)^2 \; , 
$$
$$
\frac{\prt^2 F_f}{\prt w_f\prt s_f} = - \; 
\frac{1-s_f^2}{4T} \; w_f^2 g_f \; \left( h + 2w_f g_f s_f \right) \; ,
\qquad
\frac{\prt^2 F_f}{\prt s_f^2} = \frac{1}{2}\; w_f^2 g_f \left( 1 - \; 
\frac{1-s_f^2}{4T} \; w_f^2 g_f \right) \;  .
$$

For the considered system, we have
$$
\frac{\prt^2 F}{\prt w^2} = 2u - \; \frac{1}{2}\;s_1^2 - \;
\frac{1-s_1^2}{4T} \; ( h + 2w s_1 )^2 - \;
\frac{1-s_2^2}{4T} \; h^2 \; , 
$$
$$
\frac{\prt^2 F}{\prt w\prt s_1} = - \; 
\frac{1-s_1^2}{4T} \; w^2 \; ( h + 2w s_1 ) \; ,
$$
$$
\frac{\prt^2 F}{\prt w\prt s_2} = \frac{\prt^2 F}{\prt s_1\prt s_2} =
\frac{\prt^2 F}{\prt s_2^2} = 0 \; ,
\qquad
\frac{\prt^2 F}{\prt s_1^2} = 
\frac{1}{2} \; w^2 \left( 1 - \; \frac{1-s_1^2}{2T} \; w^2 \right) \; .
$$

The minimum of the free energy implies the stability conditions that for 
the present case become
$$
\frac{\prt^2 F}{\prt w^2} > 0 \; , \qquad 
\frac{\prt^2 F}{\prt s_1^2} > 0 \; ,
$$
\be
\label{74}
\frac{\prt^2 F}{\prt w^2} \; \frac{\prt^2 F}{\prt s_1^2} -
\left( \frac{\prt^2 F}{\prt w\prt s_1} \right)^2 > 0 \; .
\ee
  
We need to solve the system of equations for the order parameter $s_1$, given 
\linebreak in (\ref{64}) and satisfying condition (\ref{55}), for the order parameter 
$s_2$, given in (\ref{66}) and satisfying condition (\ref{56}), and for the 
probability of ferromagnetic phase $w$, defined in (\ref{73}) and satisfying 
conditions (\ref{28}). If there occur several solutions, it is necessary to 
choose the solution that  corresponds to the minimal free energy and satisfies 
the stability \mbox{conditions (\ref{74})}.

Furthermore, it is necessary to choose the state with the minimal free energy 
between the free energy $F$ of the mixture, free energy $F_{fer}$ of the 
pure ferromagnetic phase  
\be
\label{75}
F_{fer} = \frac{1}{2} \; \left( u + \frac{1}{2}\; s_{fer}^2 \right) -
T \; \ln \left[ 2\cosh\left( \frac{h+s_{fer}}{2T} \right) \right] \;  ,
\ee
with the order parameter
\be
\label{76}
 s_{fer} = \tanh\left( \frac{h+s_{fer}}{2T} \right) \;  ,
\ee
and the free energy of the pure paramagnetic phase
\be
\label{77}
F_{par} = \frac{1}{2} \; u - 
T \ln \; \left[ 2 \cosh\left( \frac{h}{2T} \right) \right] \; ,
\ee
with the order parameter
\be
\label{78}
s_{par} = \tanh\left( \frac{h}{2T} \right) \;   .
\ee

\section{Results and Discussion}\label{sect:eight}

We have derived the model of a mixed system describing the coexistence of 
different phases when at least one of the phases represents nanoscale regions 
of a competing phase inside a host phase. The spatial distribution of the 
phases is random. This picture is often termed nanoscale phase separation. 
As a concrete example, we have studied the mixture of ferromagnetic and 
paramagnetic phases, modeling a ferromagnet with paramagnetic fluctuations. 
The choice of this example is dictated by the fact that spin models serve 
as typical illustrations of phase transitions of different nature. 

After averaging over phase configurations, we obtain a renormalized 
Hamiltonian, taking into account the coexistence of mesoscopic phases. In 
the resulting effective picture, thermodynamic potentials are represented 
as the sums of replicas characterizing different phases. This, however, is 
not a simple sum of the terms corresponding to pure phases, as in the case 
of the Gibbs macroscopic mixture, where, for instance, free energy is a 
linear combination, in our case
\be
\label{79}
 F_G = w_1 F_{fer} + w_2 F_{par} \;  .
\ee

The separation of phases is connected with the existence of surface free 
energy. The latter is not a microscopic notion and is not defined at the 
level of operators and microscopic states. The surface free energy is a 
thermodynamic notion defined by the difference between the actual free 
energy of the system and the free energy of the Gibbs macroscopic\linebreak mixture 
\cite{Ono_46,Rusanov_47,Kjelstrup_48}. That is, the surface free energy is 
defined by the difference
\be
\label{80}
 F_{sur}= F - F_G \; .
\ee

In our case, this is
\be
\label{81}
F_{sur} =F_1 + F_2 -w_1 F_{fer} - w_2 F_{par} \; .
\ee  

Contrary to a pure phase needing a one-order parameter (that can be a vector 
or a tensor), the mixed state requires, for its correct description, a larger 
number of parameters. Thus, compared to the pure ferromagnetic phase, 
described by a single order parameter $s_{fer}$, the mixed 
ferromagnetic--paramagnetic state needs three parameters: the order parameter 
(reduced magnetization) of the ferromagnetic component, $s_1$, the order 
parameter (reduced magnetization) of the paramagnetic component, $s_2$, and 
the probability of one of the phases, say $w$, with the probability of the 
other phase given by $1-w$.

\begin{figure}[ht]
\centerline{
\hbox{ \includegraphics[width=8cm]{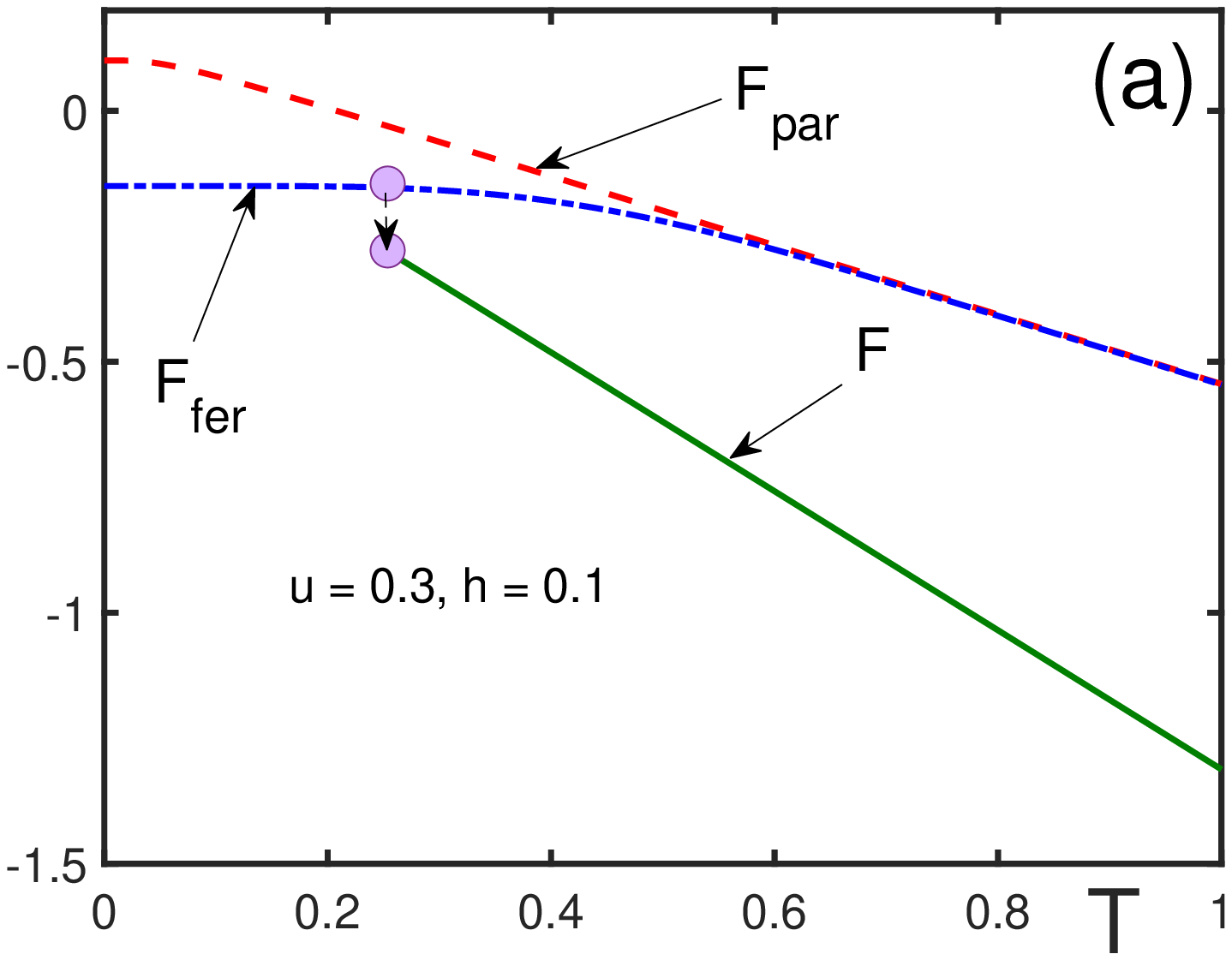} \hspace{1cm}
\includegraphics[width=8cm]{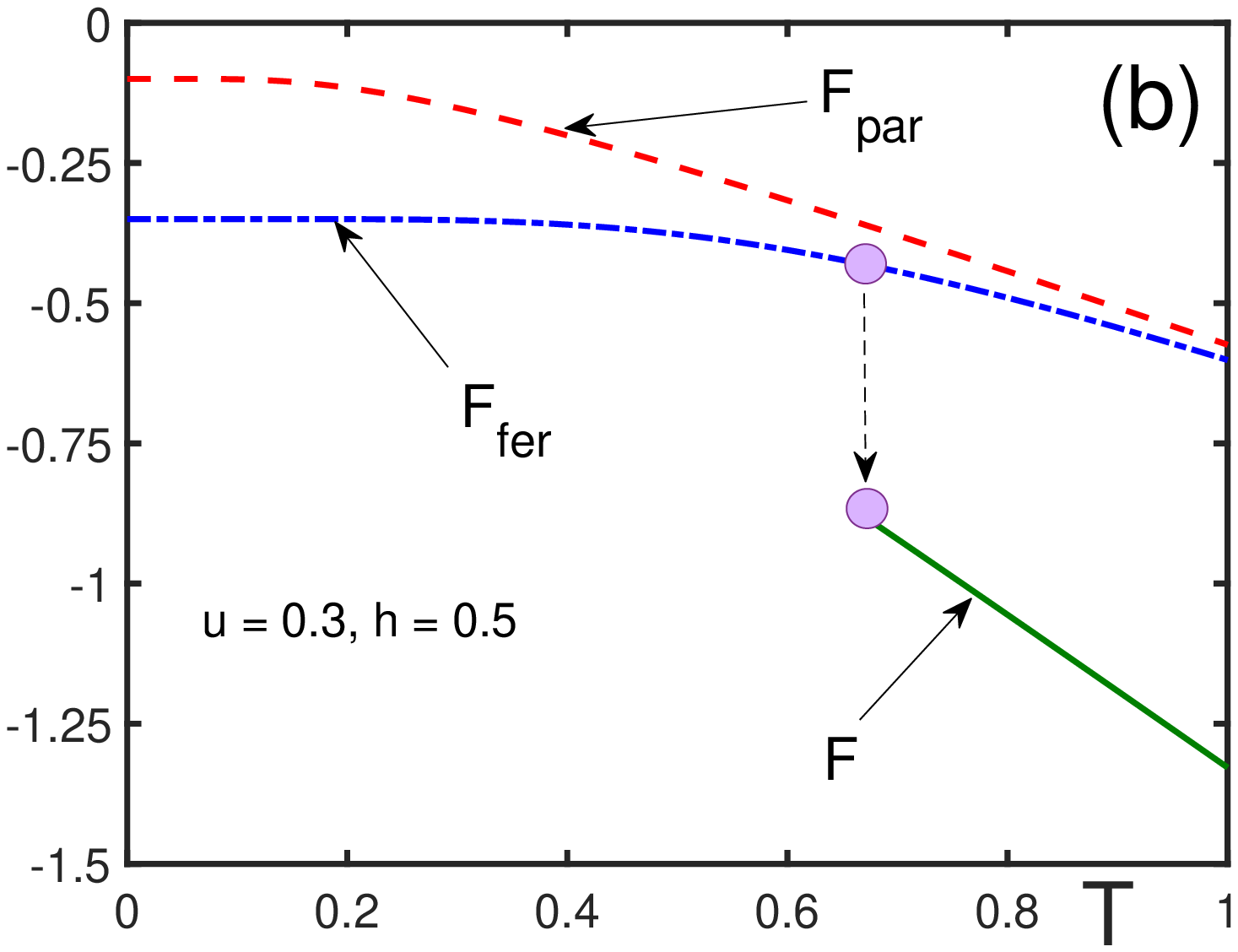}  } }
\caption{Free energies of the mixed state, $F$ (solid line), 
ferromagnetic state, $F_{fer}$ (dash--dotted line), and  of the paramagnetic 
state, $F_{par}$ (dashed line), for $u = 0.3$ and different magnetic fields: \\
(\textbf{a}) $h = 0.1$; (\textbf{b}) $h = 0.5$.
}
\label{fig:Fig.1}
\end{figure}

\begin{figure}[ht]
\centerline{
\hbox{ \includegraphics[width=8cm]{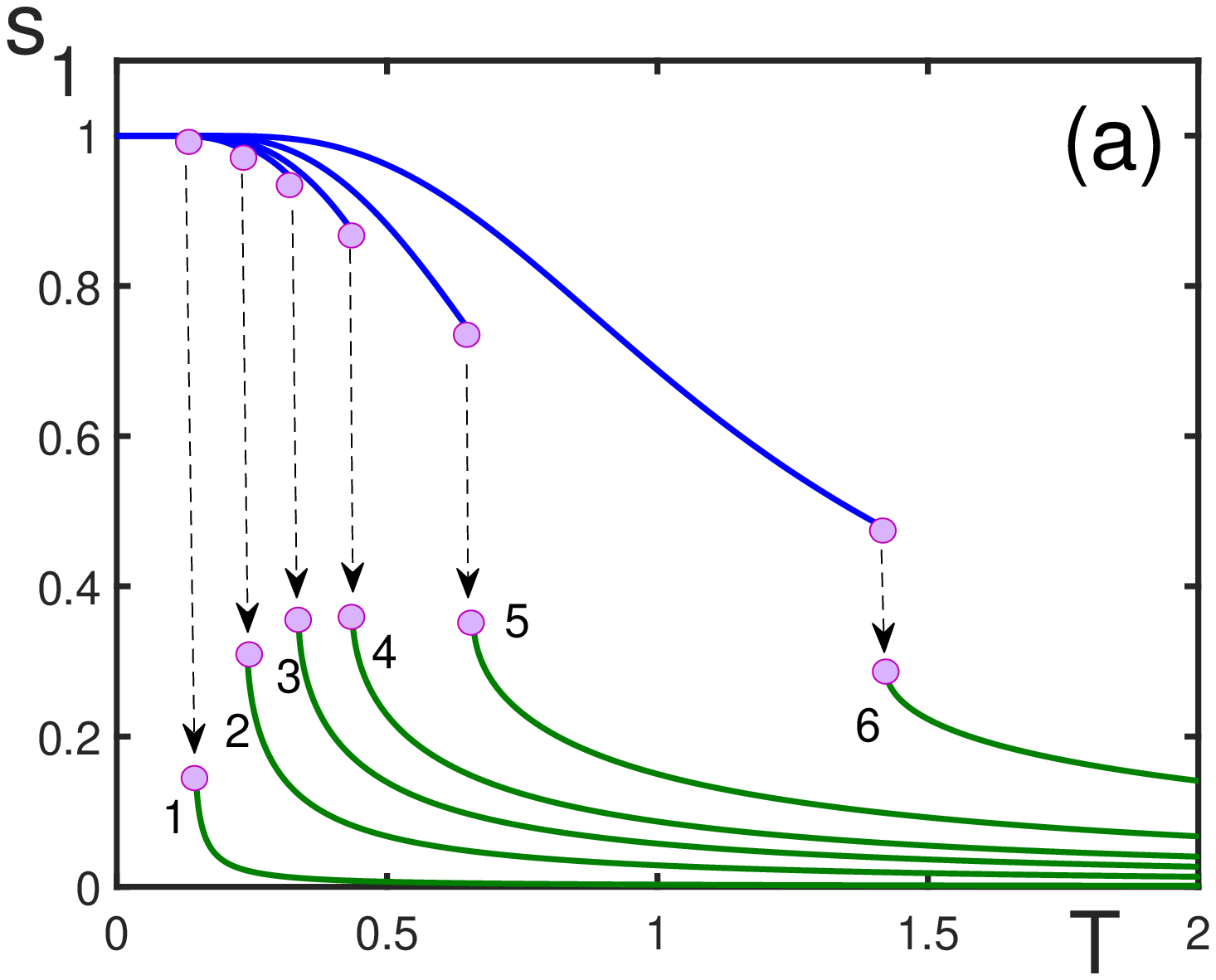} \hspace{1cm}
\includegraphics[width=8cm]{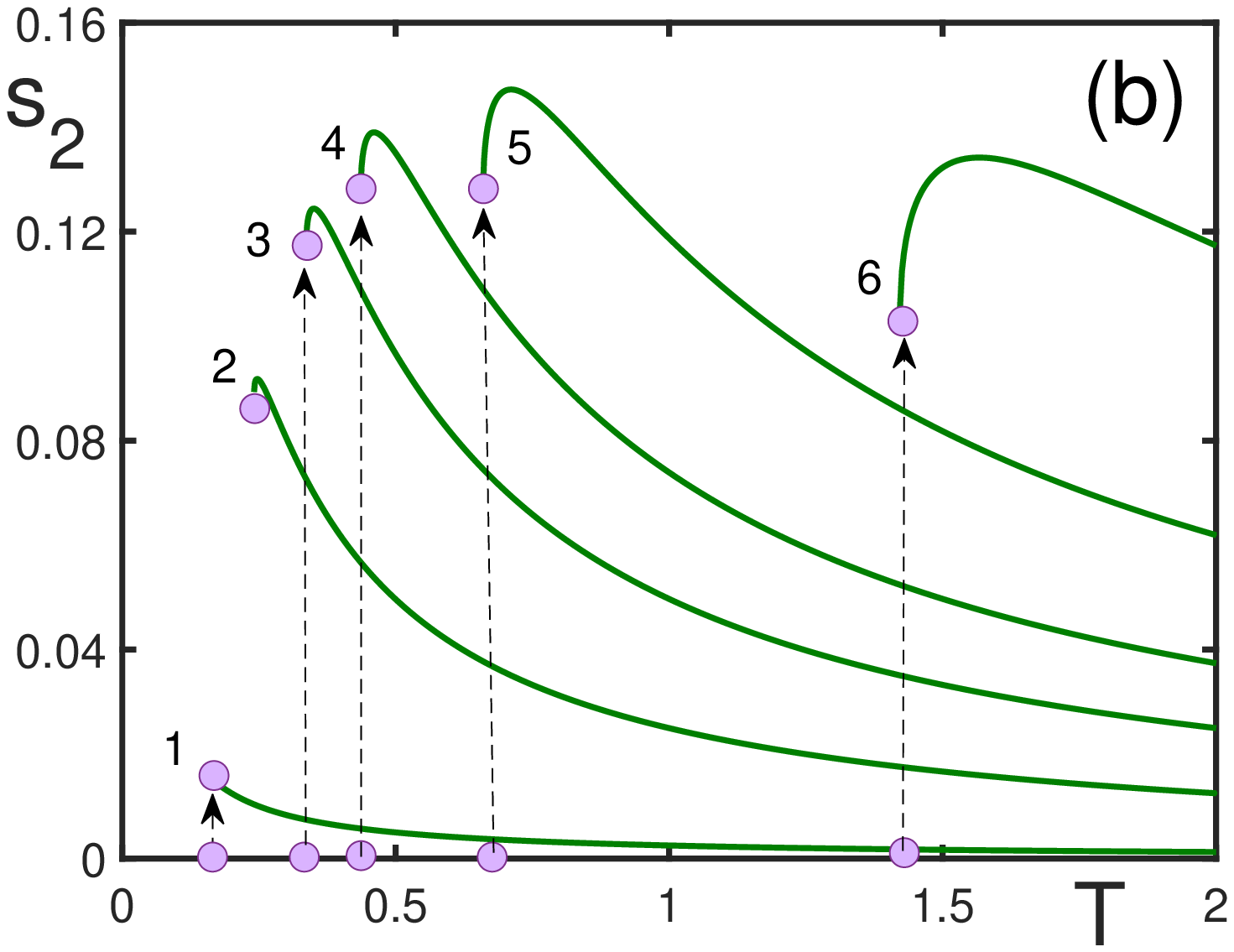}  } }
\vspace{12pt}
\centerline{
\hbox{ \includegraphics[width=8cm]{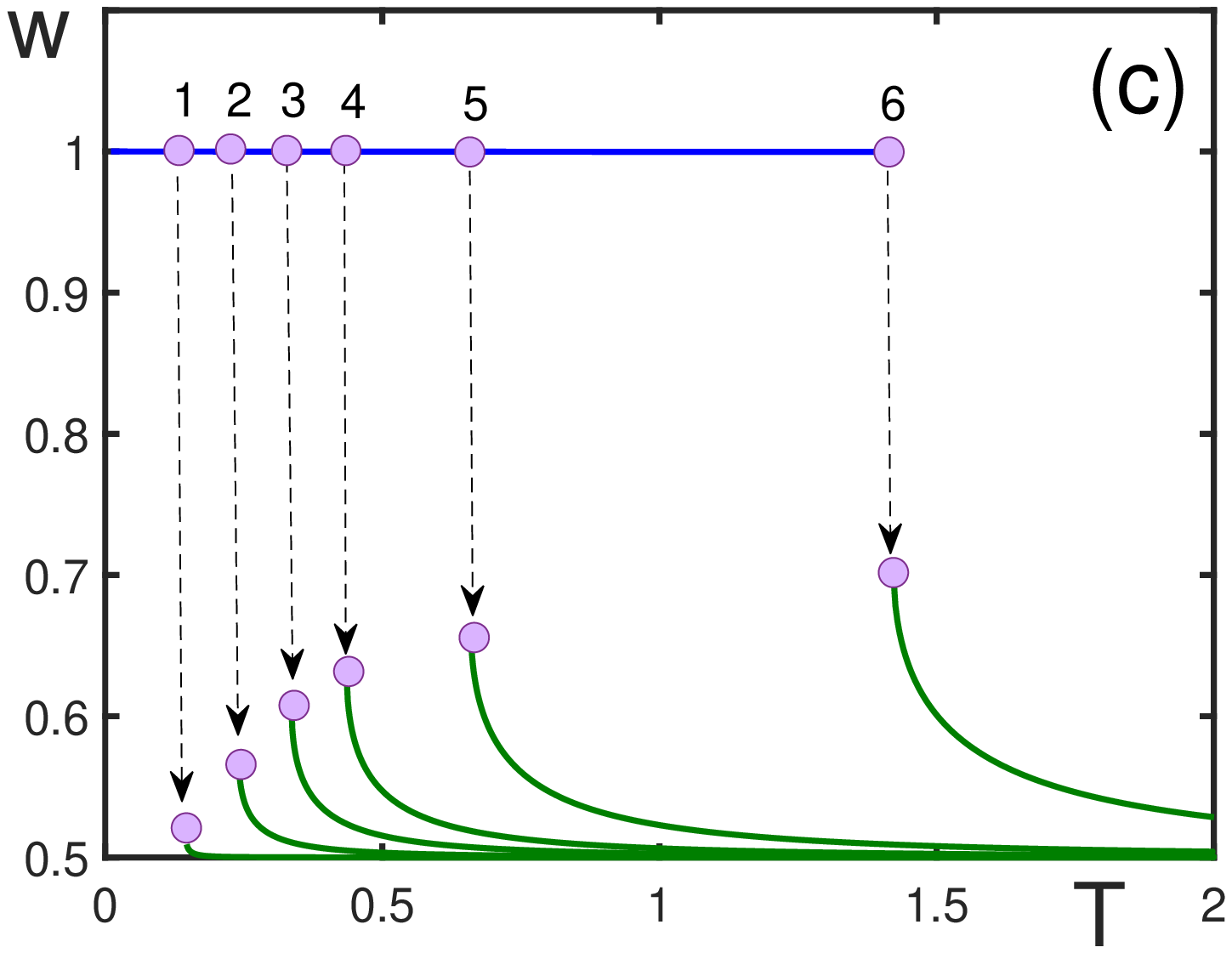}   } }
\caption{Order parameters $s_1$, $s_2$, and $w$ as functions of 
dimensionless temperature $T$, for $u = 0.3$ and different fields: \\
(1) $h = 0.01$; (2) $h = 0.1$; (3) $h = 0.2$; (4) $h = 0.3$; (5) $h = 0.5$; 
(6) $h = 1$. 
The corresponding nucleation temperatures are: \\
(1) $T_n = 0.15$; (2) $T_n = 0.24$; (3) $T_n = 0.34$; (4) $T_n = 0.44$; 
(5) $T_n = 0.66$; (6) $T_n = 1.42$. 
}
\label{fig:Fig.2}
\end{figure}

\begin{figure}[ht]
\centerline{
\hbox{ \includegraphics[width=8cm]{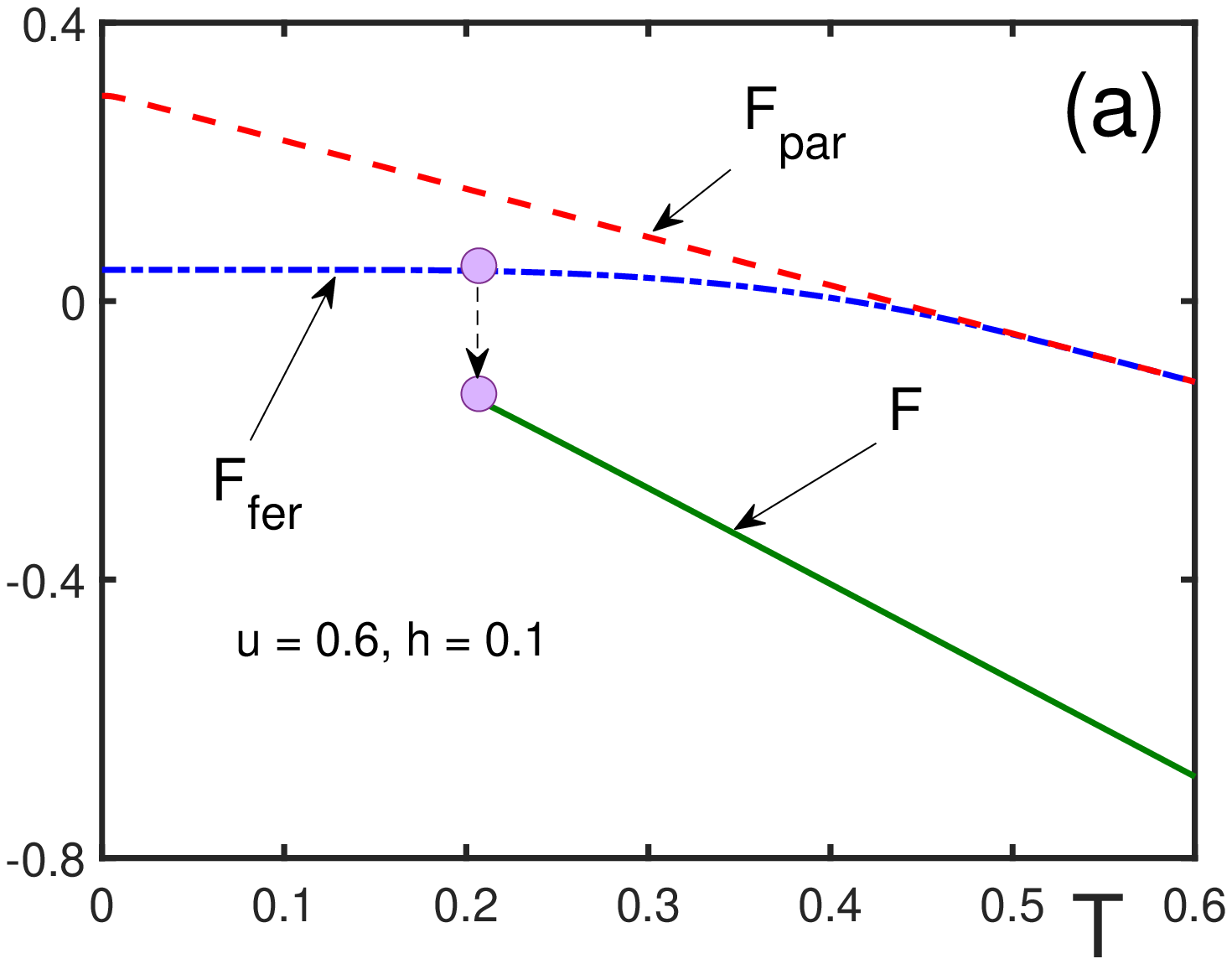} \hspace{1cm}
\includegraphics[width=8cm]{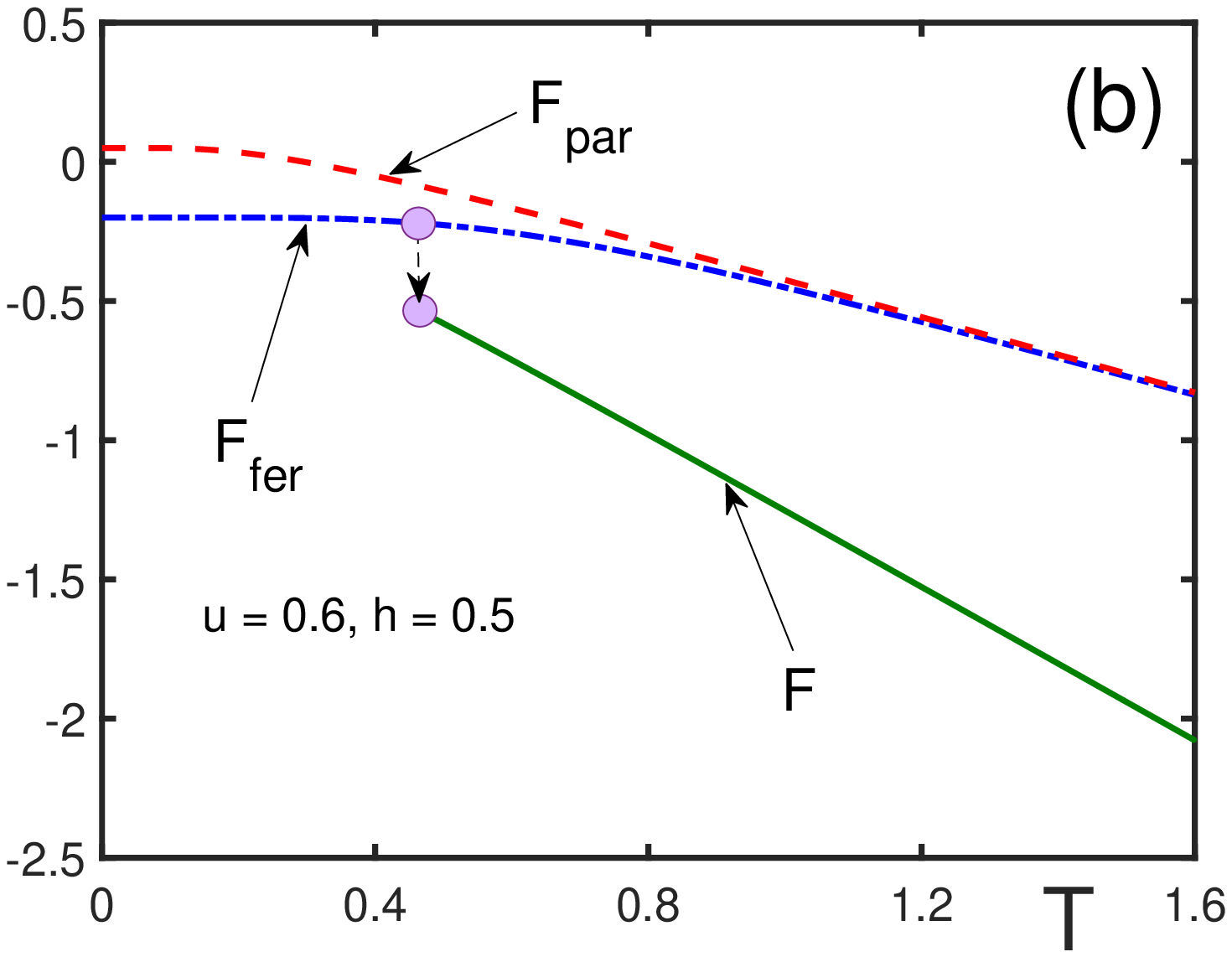}  } }
\caption{Free energies of the mixed state, $F$ (solid line), 
ferromagnetic state, $F_{fer}$ (dash-dotted line), and  of the paramagnetic 
state, $F_{par}$ (dashed line), for $u = 0.6$ and different magnetic 
fields: (\textbf{a}) $h = 0.1$; (\textbf{b}) $h = 0.5$.
}
\label{fig:Fig.3}
\end{figure}

\begin{figure}[ht]
\centerline{
\hbox{ \includegraphics[width=8cm]{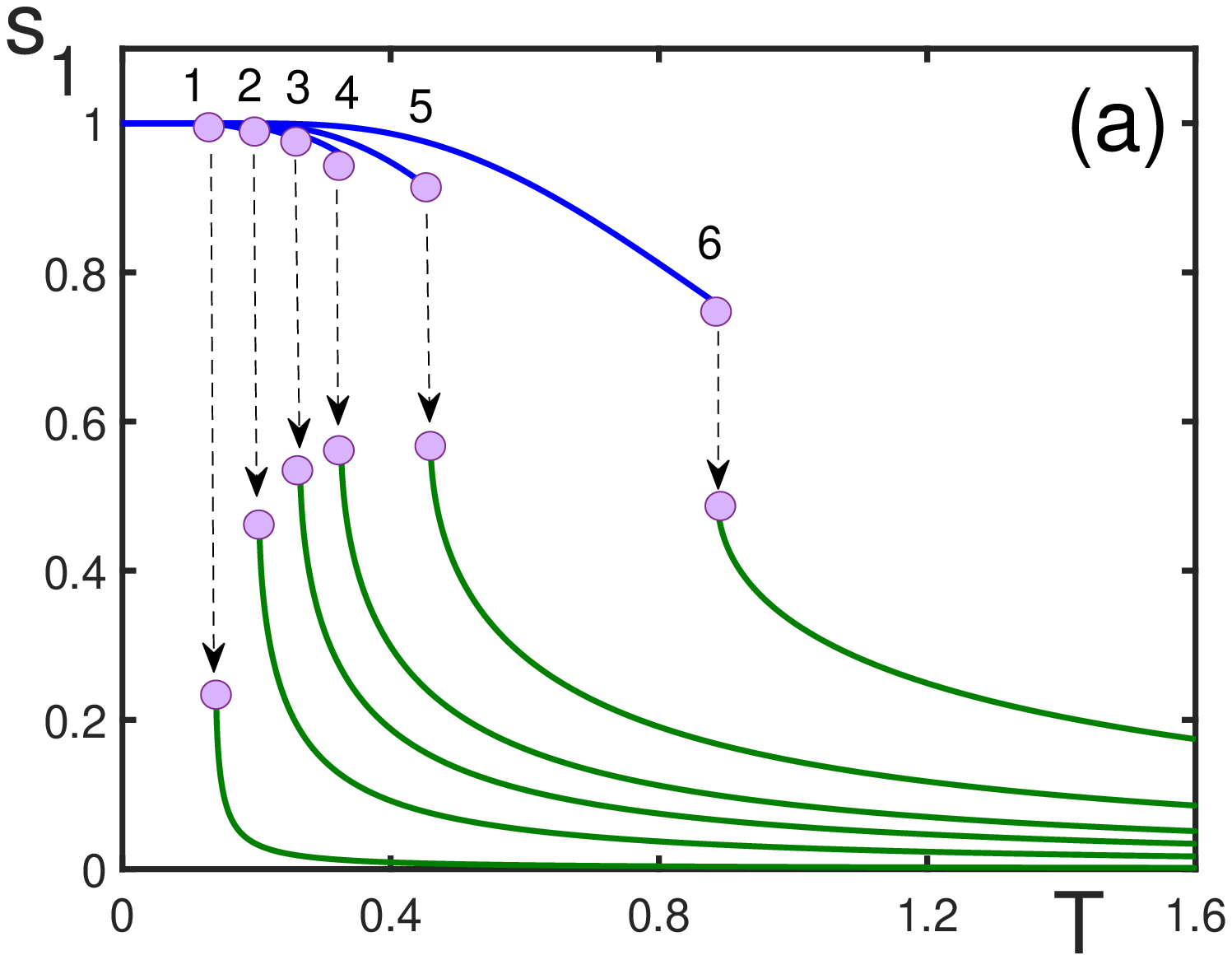} \hspace{1cm}
\includegraphics[width=8cm]{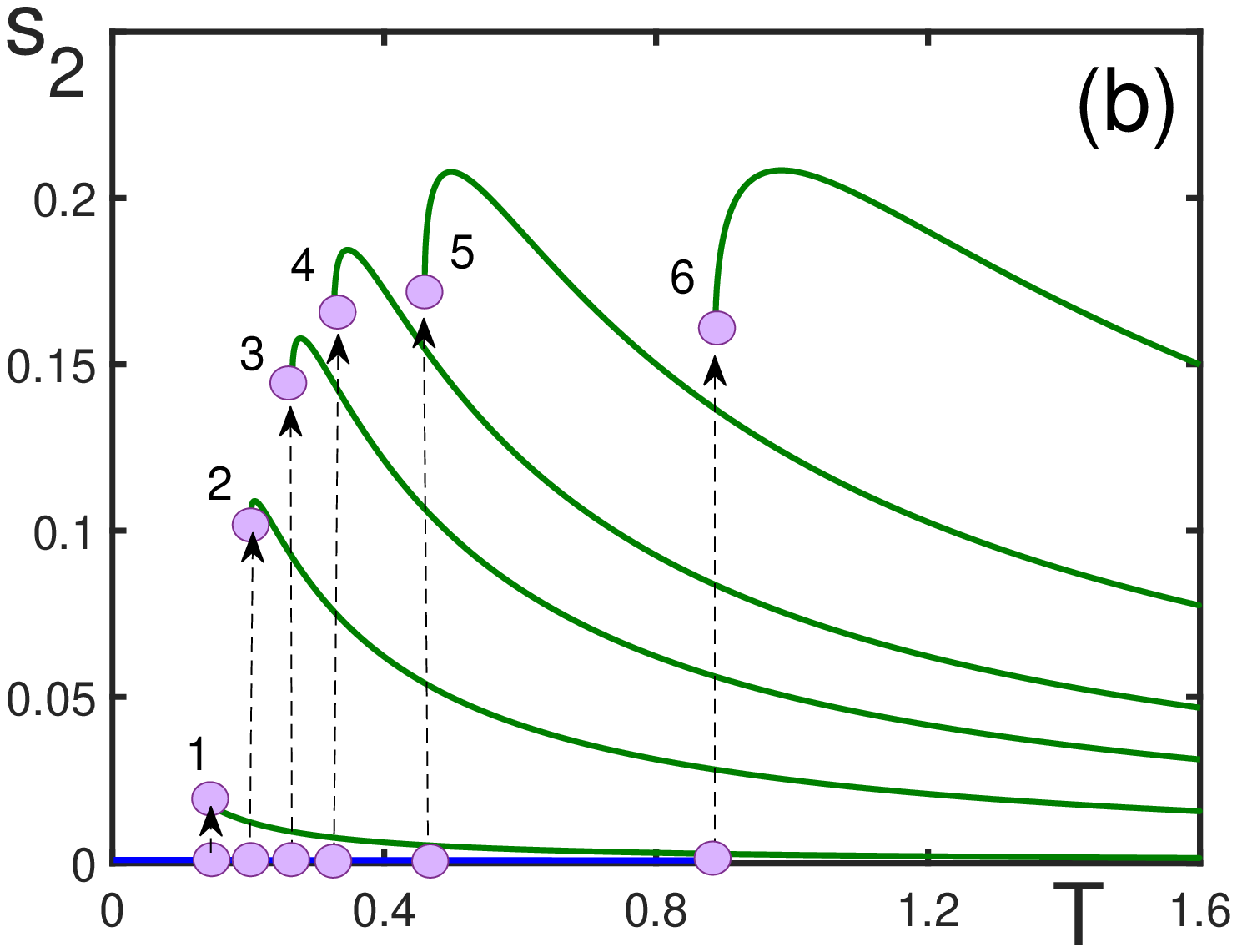}  } }
\vspace{12pt}
\centerline{
\hbox{ \includegraphics[width=8cm]{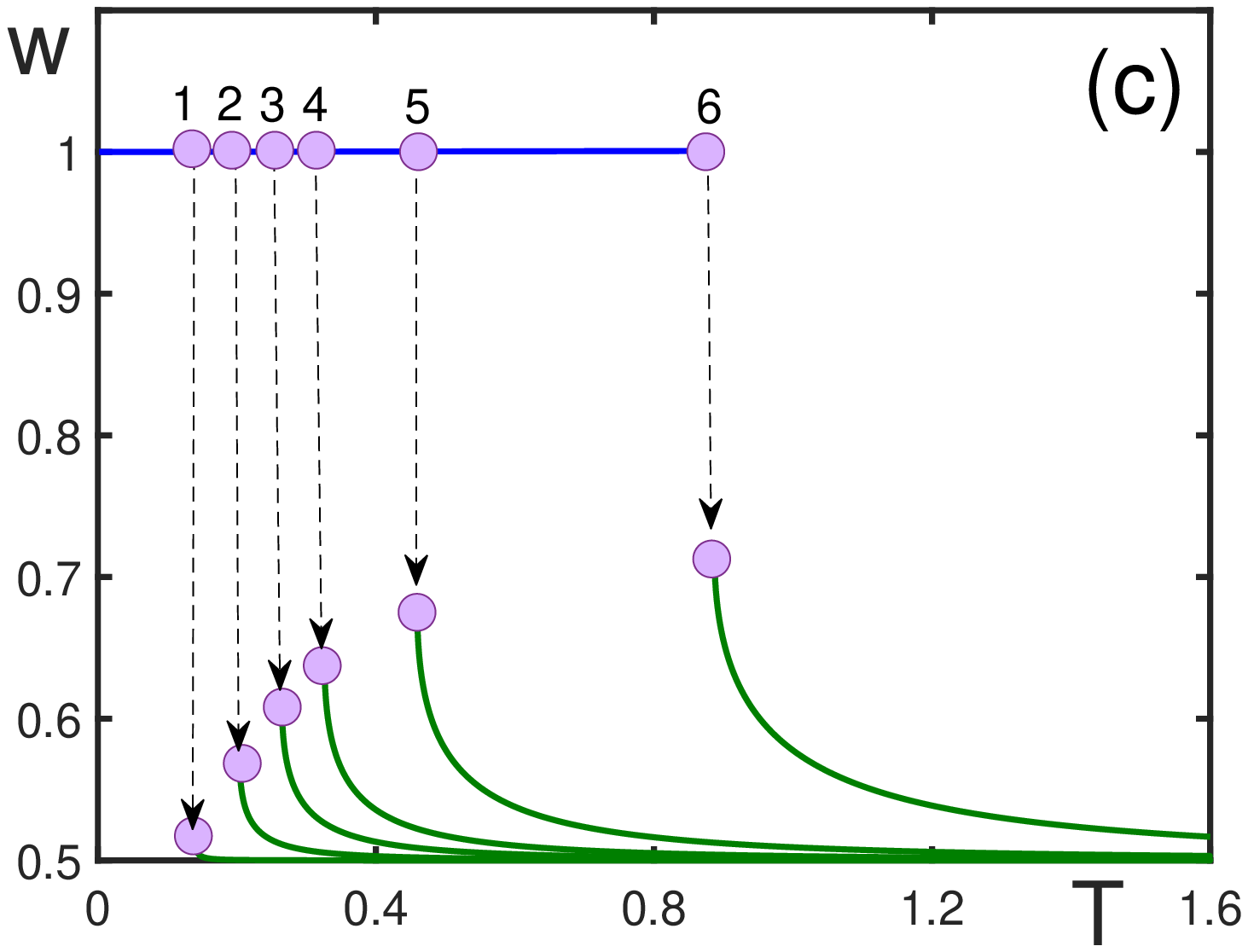}   } }
\caption{Order parameters $s_1$, $s_2$, and $w$ as functions of 
dimensionless temperature $T$, for $u = 0.6$ and different fields: \\
(1) $h = 0.01$; (2) $h = 0.1$; (3) $h = 0.2$; (4) $h = 0.3$; (5) $h = 0.5$; 
(6) $h = 1$. The corresponding nucleation temperatures are: %\linebreak
(1) $T_n = 0.14$; (2) $T_n = 0.20$; (3) $T_n = 0.26$; (4) $T_n = 0.33$; 
(5) $T_n = 0.46$; (6) $T_n = 0.89$.
}
\label{fig:Fig.4}
\end{figure}

\begin{figure}[ht]
\centerline{
\hbox{ \includegraphics[width=8cm]{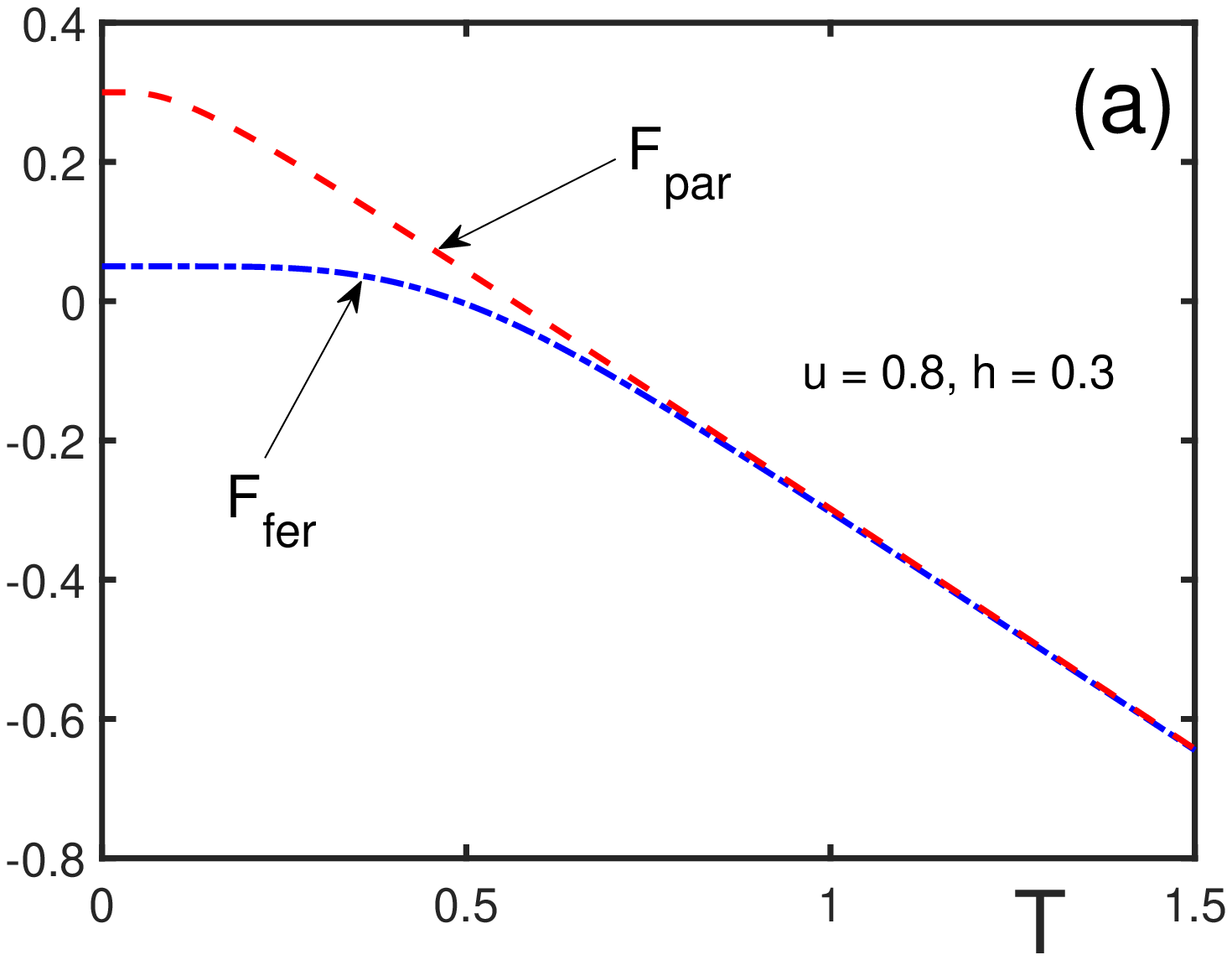} \hspace{1cm}
\includegraphics[width=8cm]{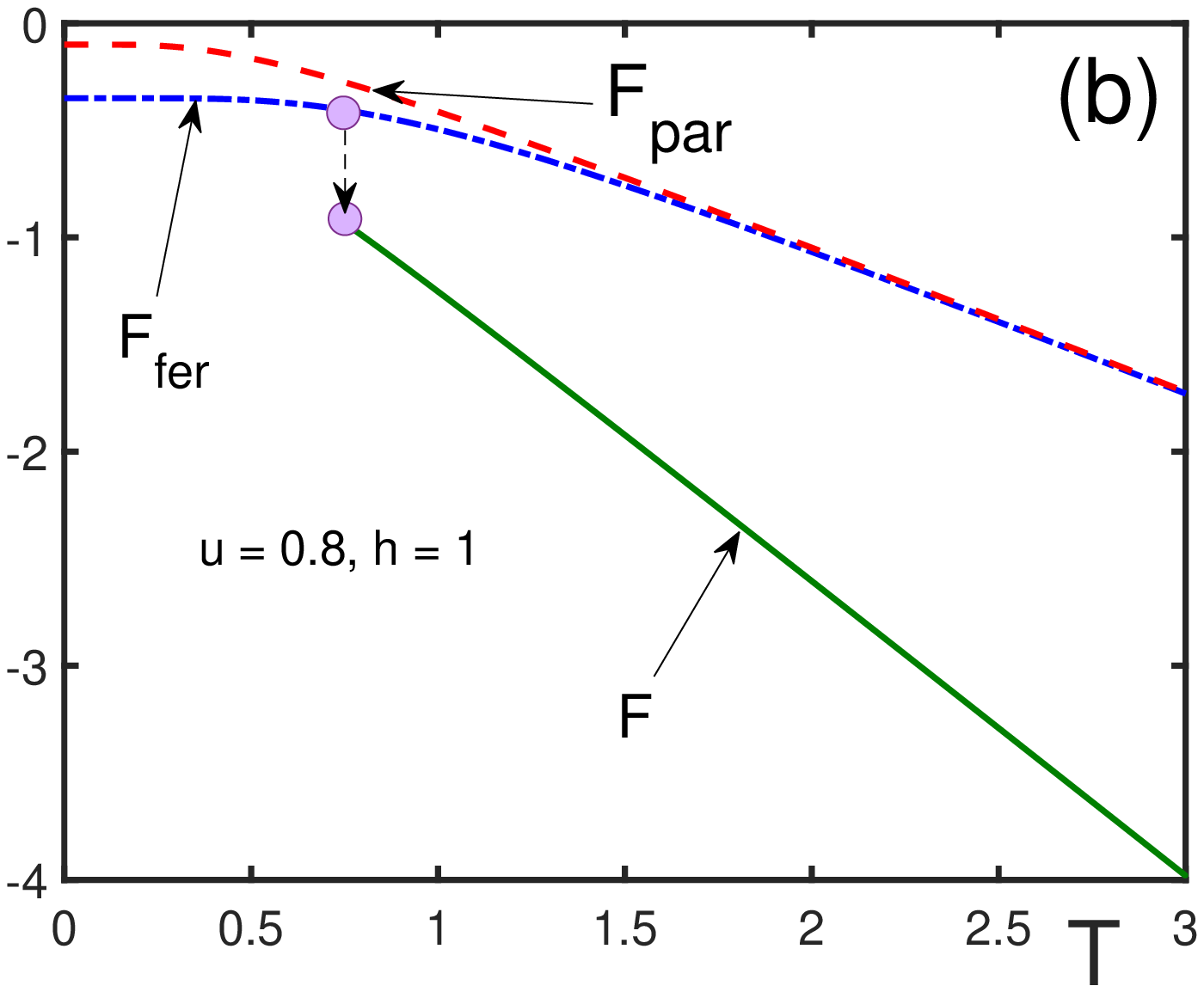}  } }
\caption{Free energies of the mixed state, $F$ (solid line), 
ferromagnetic state, $F_{fer}$ (dash--dotted line), and  of the paramagnetic 
state, $F_{par}$ (dashed line), for $u = 0.8$ and different magnetic fields: \\
(\textbf{a}) $h = 0.3$; (\textbf{b}) $h = 1$. For $h = 0.3$, the mixed state 
is not stable. For $h = 1$, the zeroth-order nucleation transition occurs at 
the nucleation temperature $T_n = 0.72$.
}
\label{fig:Fig.5}
\end{figure}

\begin{figure}[ht]
\centerline{
\hbox{ \includegraphics[width=8cm]{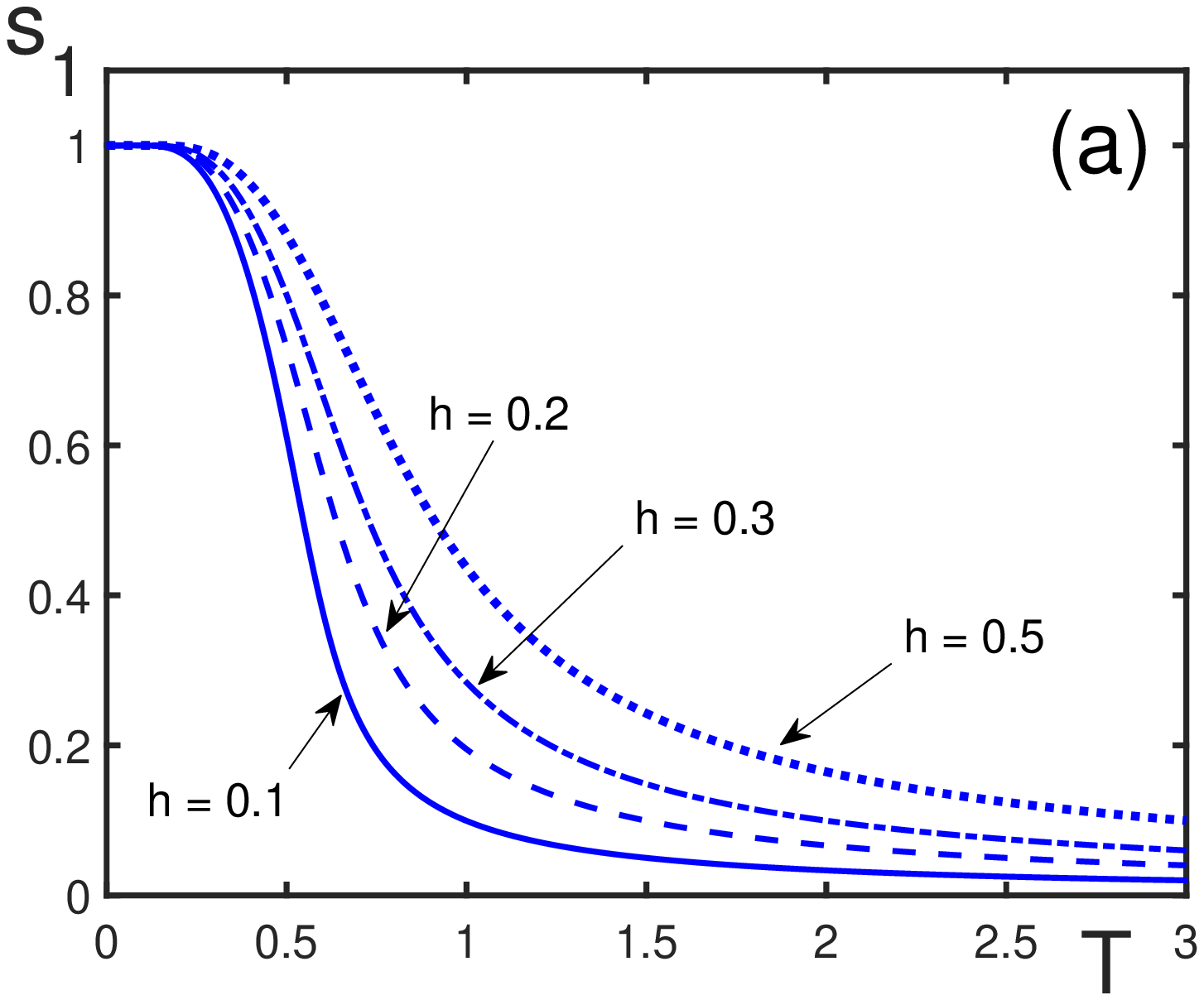} \hspace{1cm}
\includegraphics[width=8cm]{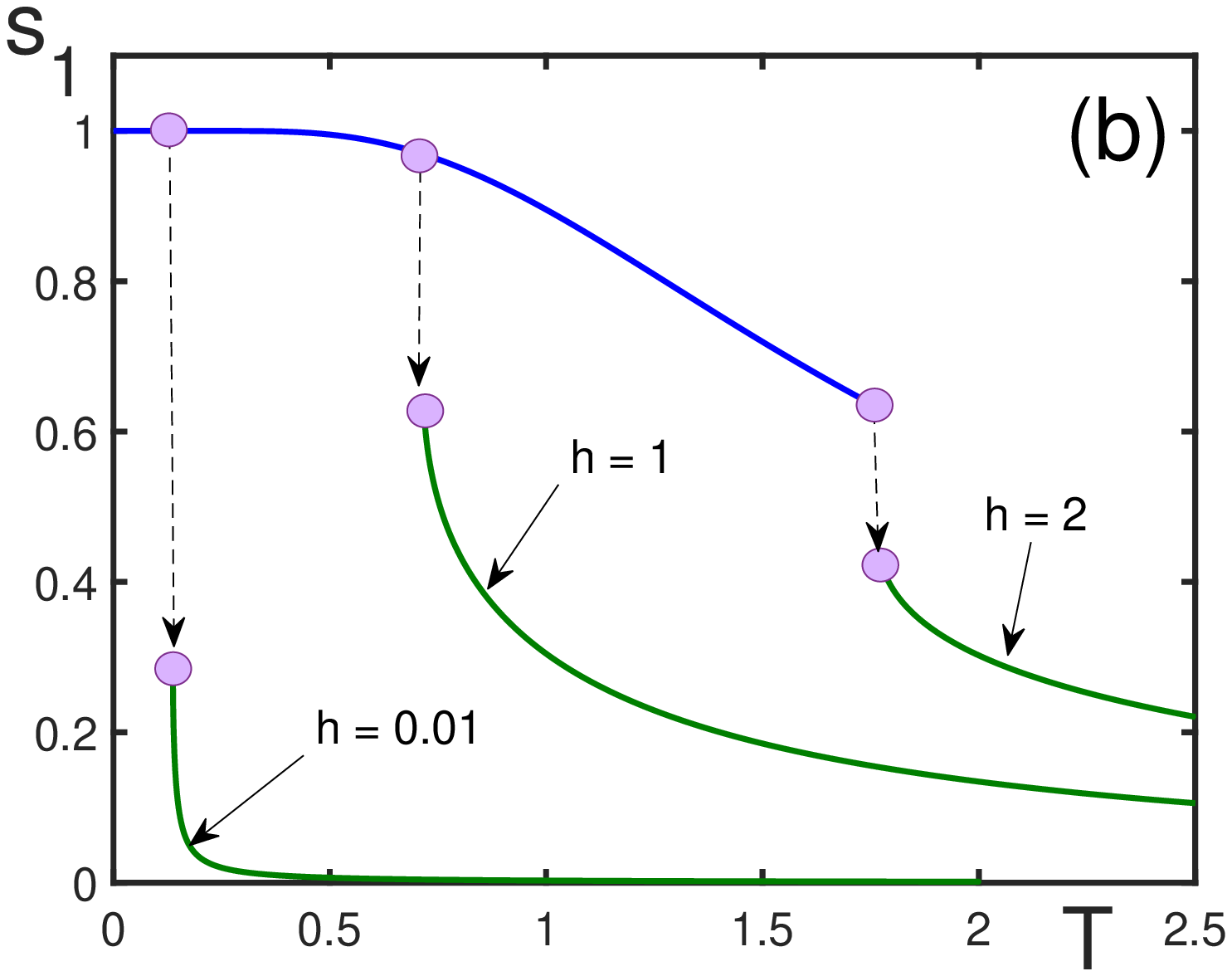}  } }
\vspace{12pt}
\centerline{
\hbox{ \includegraphics[width=8cm]{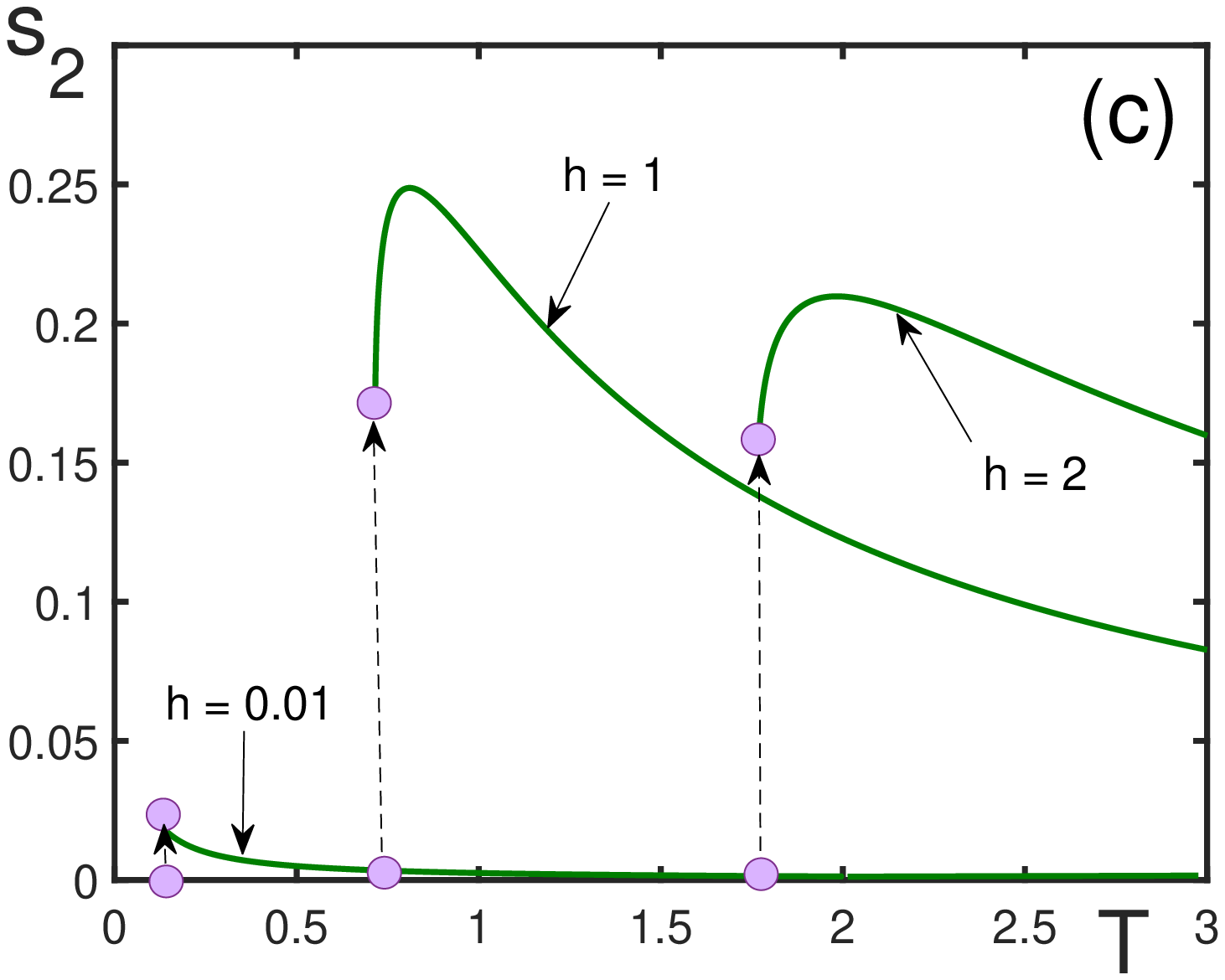} \hspace{1cm}
\includegraphics[width=8cm]{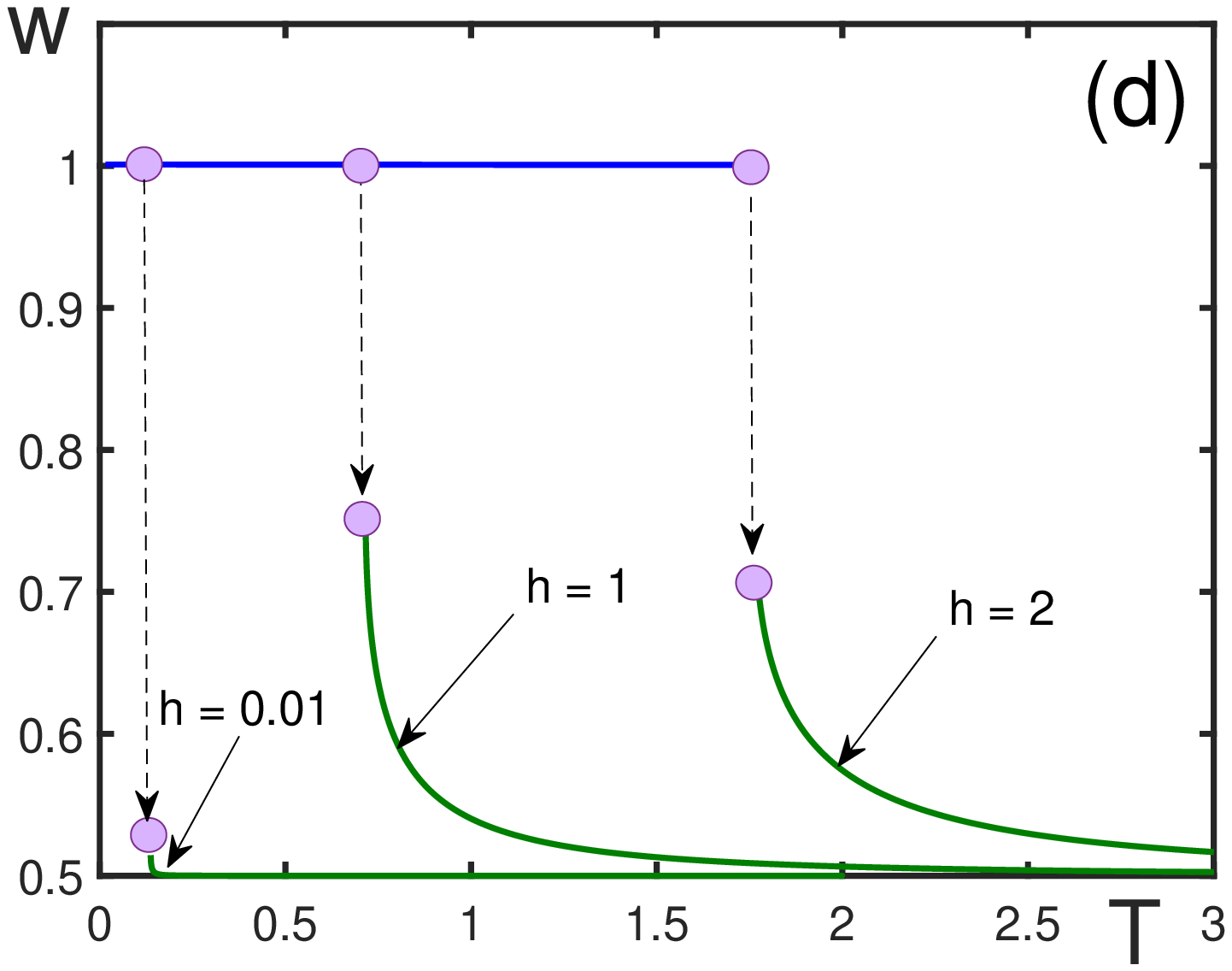}  } }
\caption{Order parameters $s_1$, $s_2$, and $w$ as functions 
of dimensionless temperature $T$, for $u = 0.8$ and different fields: \\
(\textbf{a}) $h = 0.1$ (solid line), $h = 0.2$ (dashed line); $h = 0.3$ (dash--dotted 
line); $h = 0.5$ (dotted line); (\textbf{b}) $h = 0.01$; $h = 1$; $h = 2$. The 
corresponding nucleation temperatures are $T_n = 0.14$, $T_n = 0.72$, 
and $T_n = 1.77$; (\textbf{c}) $h = 0.01$; $h = 1$; $h = 2$; 
(\textbf{d}) $h = 0.01$; $h = 1$; $h = 2$.
}
\label{fig:Fig.6}
\end{figure}

\begin{figure}[ht]
\centerline{
\hbox{ \includegraphics[width=8cm]{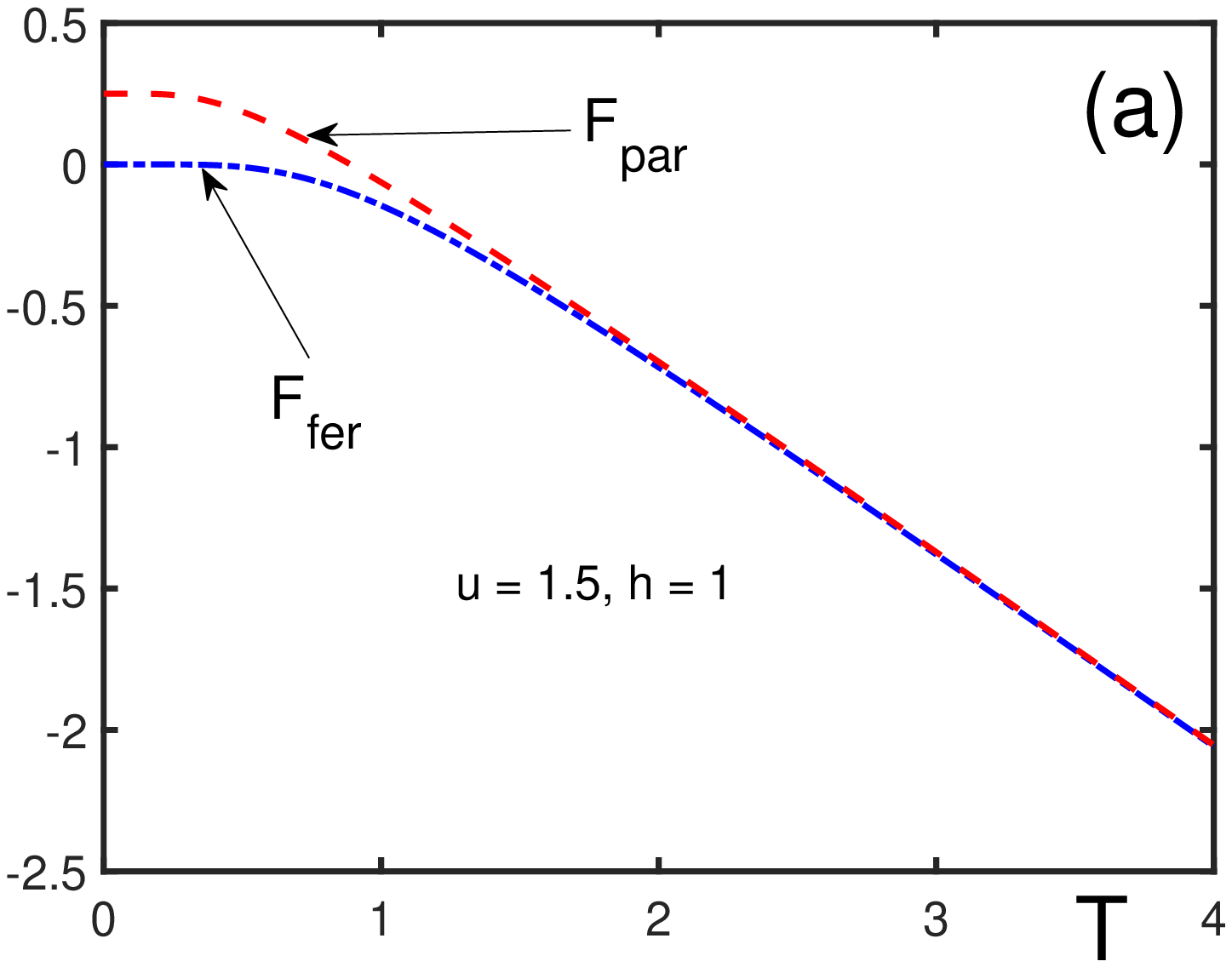} \hspace{1cm}
\includegraphics[width=8cm]{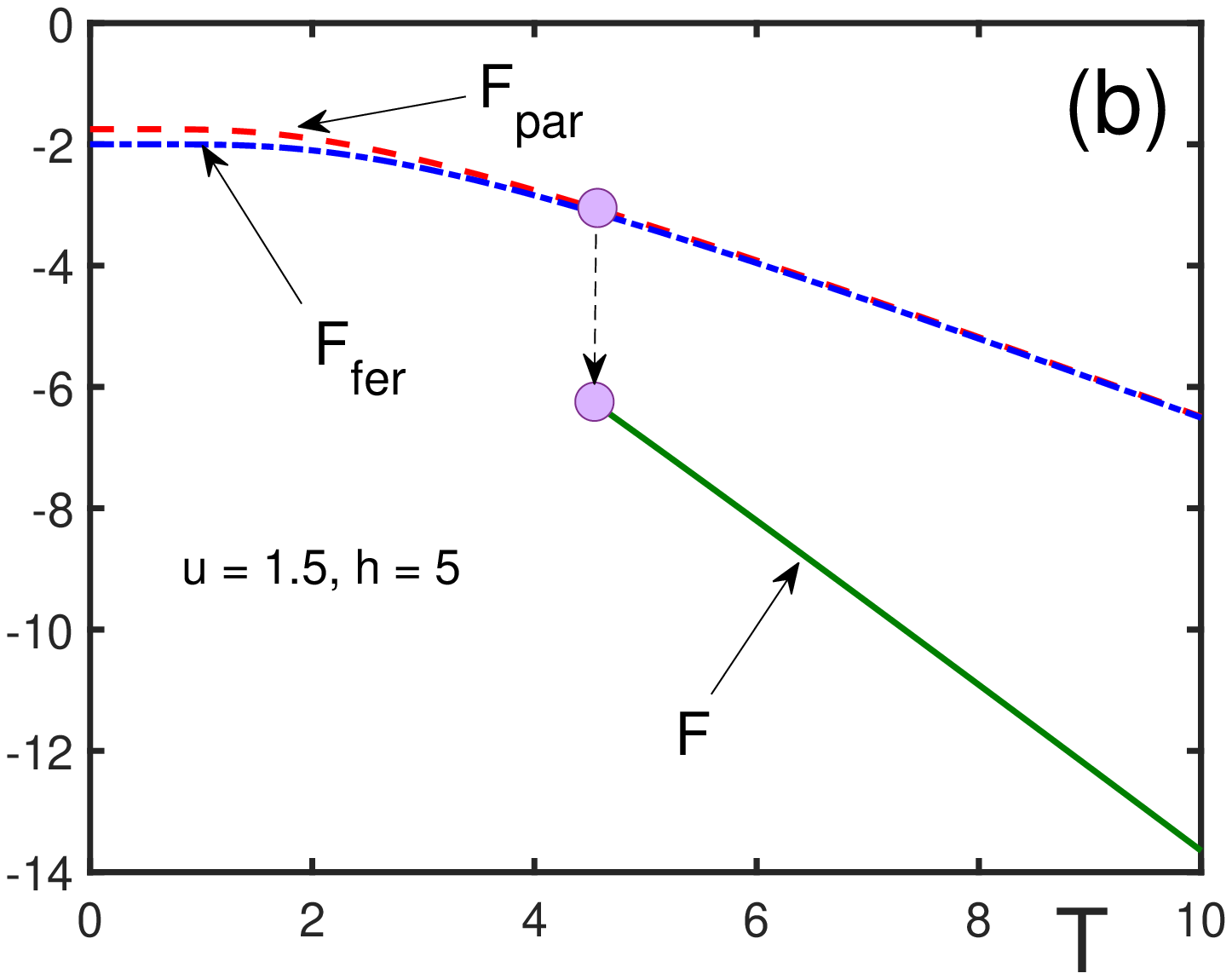}  } }
\caption{Free energies of the mixed state, $F$ (solid line), 
ferromagnetic state, $F_{fer}$ (dash--dotted line), and  of the paramagnetic 
state, $F_{par}$ (dashed line), for $u = 1.5$ and different magnetic fields: \\
(\textbf{a}) $h = 1$; (\textbf{b}) $h = 5$. For $h = 1$, the mixed state is not stable. 
For $h = 5$, the zeroth-order nucleation transition occurs at the nucleation 
temperature $T_n = 4.55$.
}
\label{fig:Fig.7}
\end{figure}

\begin{figure}[ht]
\centerline{
\hbox{ \includegraphics[width=8cm]{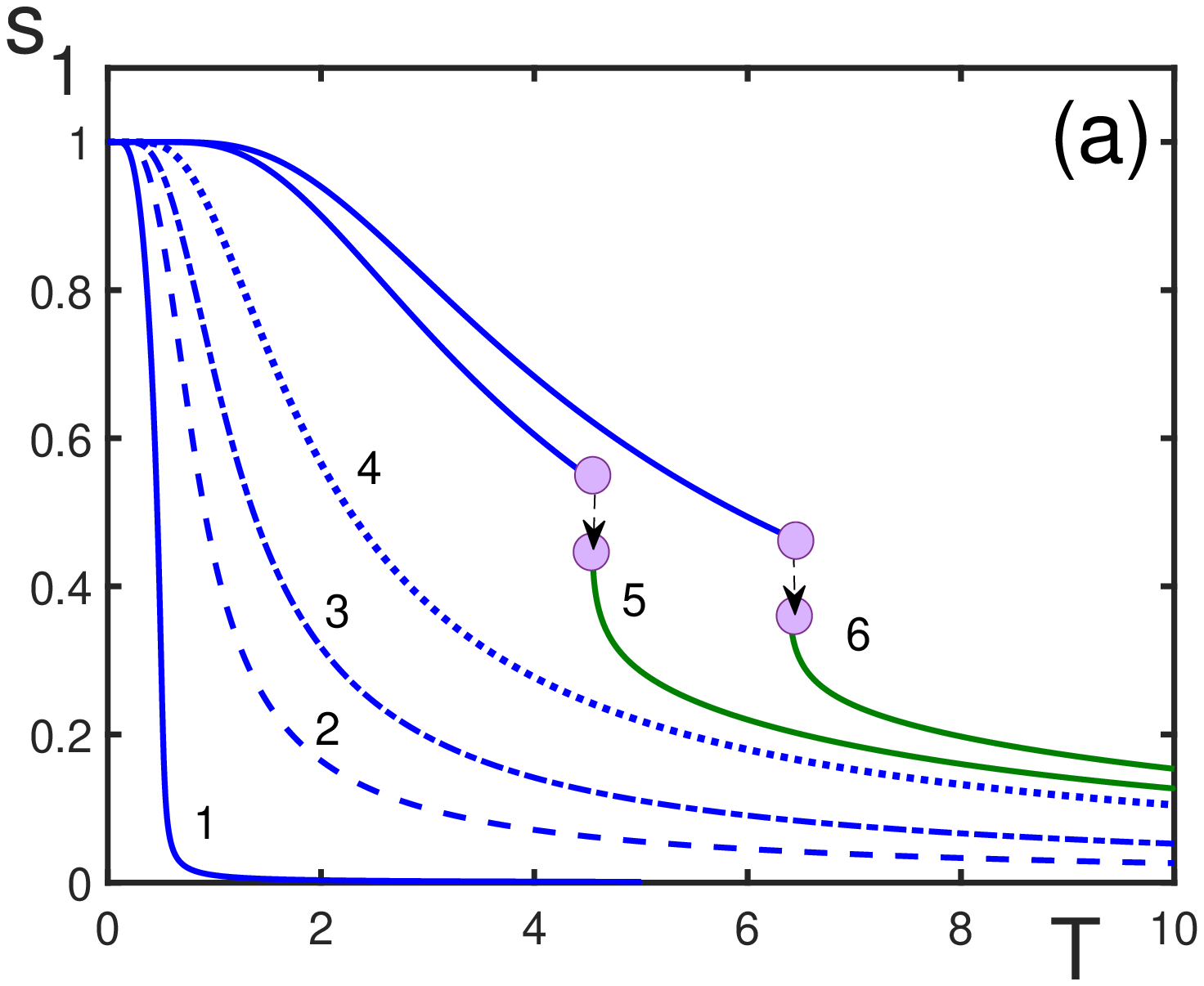} \hspace{1cm}
\includegraphics[width=8cm]{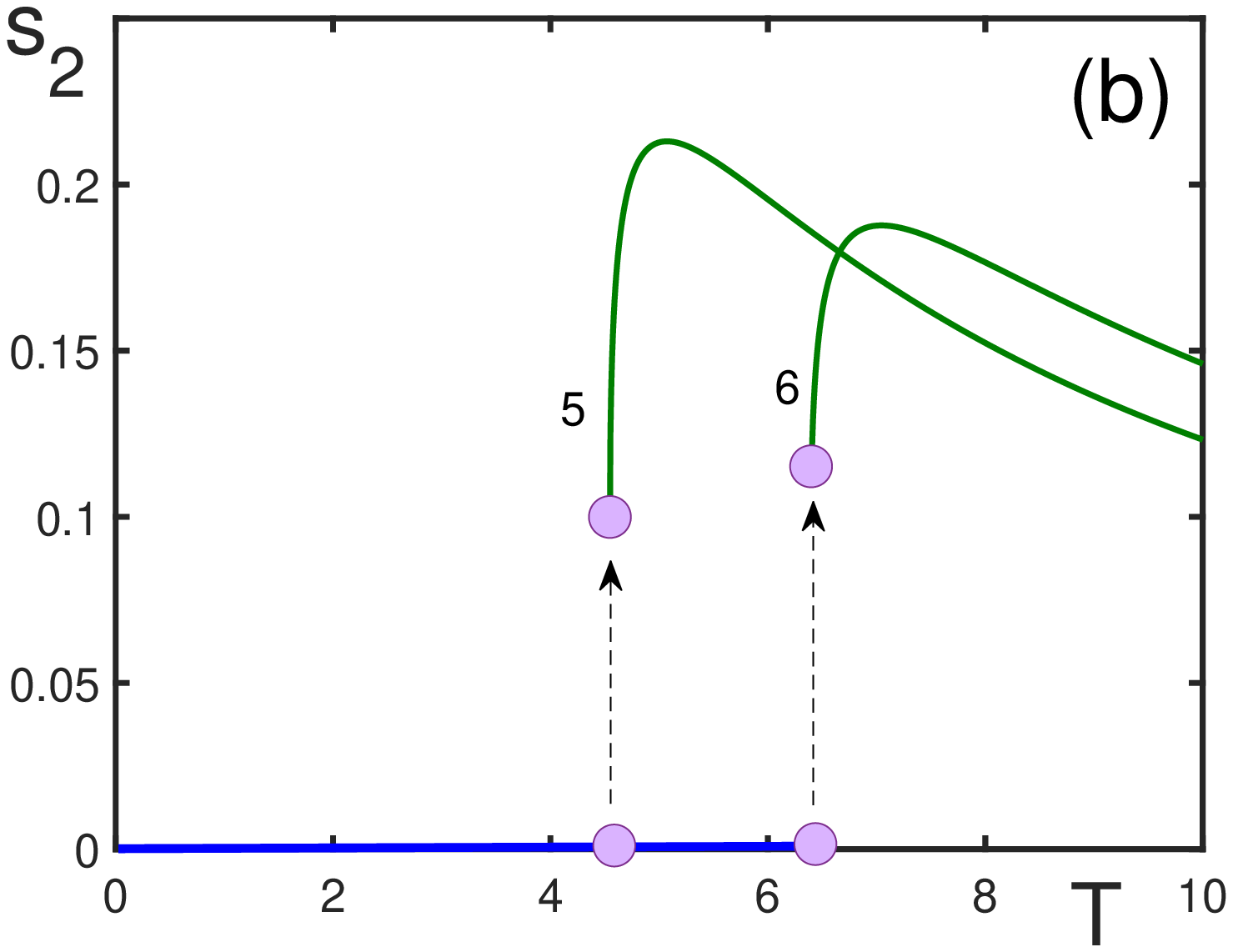}  } }
\vspace{12pt}
\centerline{
\hbox{ \includegraphics[width=8cm]{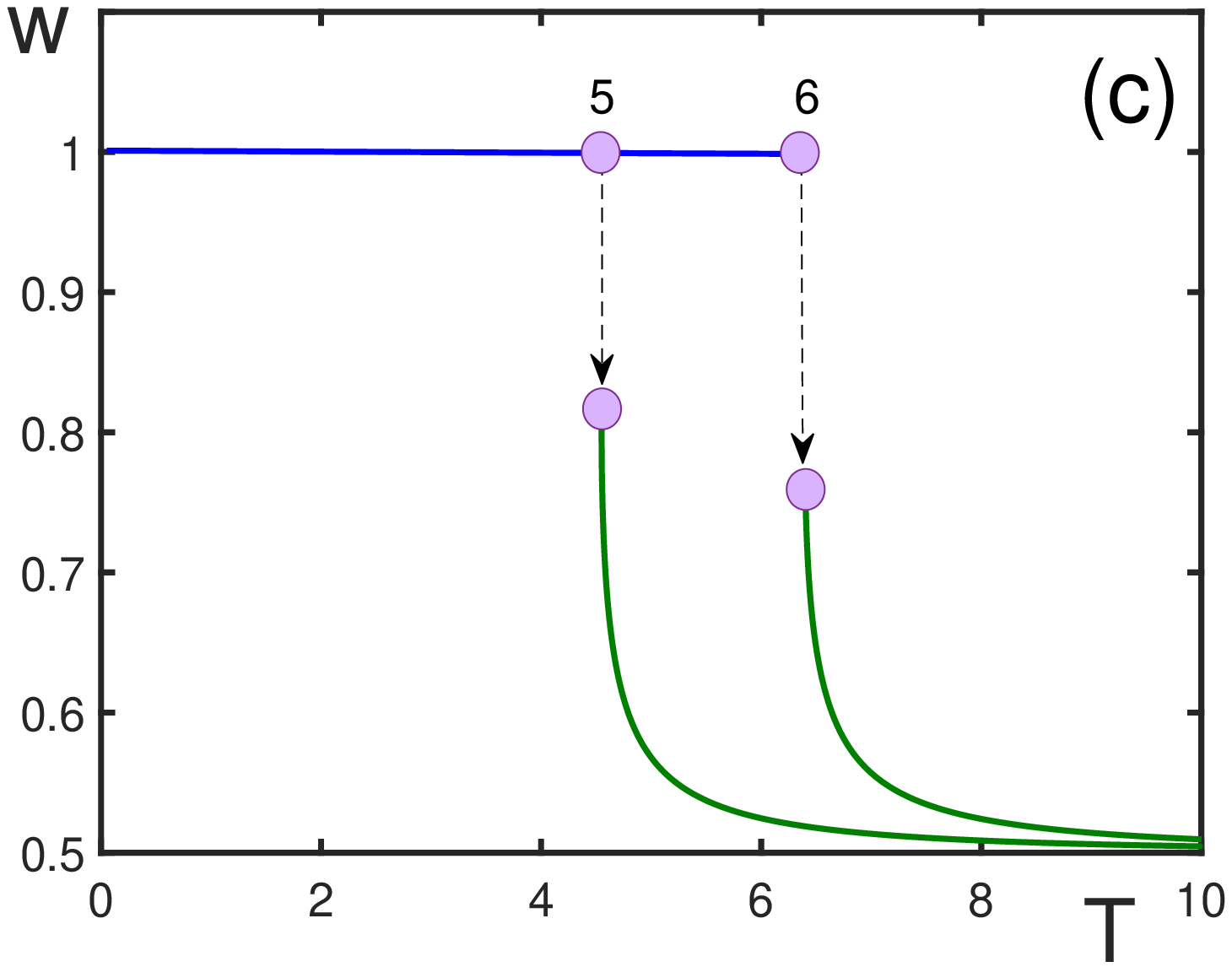}   } }
\caption{Order parameters $s_1$, $s_2$, and $w$ as functions of 
dimensionless temperature $T$, for $u = 1.5$ and different fields: \\
(1) $h = 0.01$; (2) $h = 0.5$; (3) $h = 1$; (4) $h = 2$; (5) $h = 5$; 
(6) $h = 6$. The nucleation temperatures are $T_n = 4.55$ for $h = 5$ 
and $T_n = 6.4$ for $h = 6$. 
}
\label{fig:Fig.8}
\end{figure}

In Figures \ref{fig:Fig.1}--\ref{fig:Fig.8}, we present the results of the numerical 
investigation for different parameters $u$ and $h$. Only stable solutions are shown. 
The absence of $F$ in a figure implies that $F$ is unstable. Depending on the values of 
the parameters, there can exist two types \mbox{of behavior}. 

\begin{enumerate}
\item[(i)]	At low temperatures, the system is a pure ferromagnet described by the 
free energy $F_{fer}$ and the order parameter $s_{fer} \equiv s_1$, with 
$w \equiv 1$. When increasing temperature, $F_{fer}$ gradually approaches 
$F_{par}$ corresponding to a paramagnet. The order parameter 
$s_{fer}\equiv s_1$ has the form typical of the ferromagnetic magnetization. 
This behavior, for instance, happens for $u < 0.25$ and all $h > 0$.
\item[(ii)]	For $u>0.25$, at low temperatures, below the nucleation temperature 
$T_n$, the system is a pure ferromagnet, with the free energy $F_{fer}$, 
the order parameter $s_{fer}\equiv s_1$, and $w\equiv 1$. At the nucleation 
temperature $T_n$, there appears a solution for the mixed state with the 
free energy $F$ and the order parameters $s_1$ and $s_2$. The free energy 
$F$ is lower than $F_{fer}$, but does not intersect it so that the 
nucleation is to be classified as a zeroth-order transition. 
\end{enumerate}

In this way, the nucleation transition is the transition of a system from 
a pure phase into a mixed phase. In the considered case, this is the 
transition between the pure ferromagnetic phase and a mixed state, where 
ferromagnetic regions start coexisting with \mbox{paramagnetic fluctuations}. 

As follows from the figures, the zeroth-order nucleation transition is 
accompanied by the abrupt appearance inside the ferromagnetic phase of a finite 
concentration of nanoscale paramagnetic regions. Hence, when the concentration 
of the paramagnetic admixture does not continuously grow from zero but increases 
by a jump, this suggests the possible occurrence of a zeroth-order nucleation 
transition. The appearance of paramagnetic regions can be noticed by means of 
M\"{o}ssbauer experiments.    

We show, numerically, that the nucleation transition can be of zeroth order. From 
one side, this could be a consequence of approximations involved in the process of 
calculations. From the other side, strictly speaking, nucleation is not a 
typical phase transition, because of which, it is not compulsorily required 
to be classified as %Please check intended meaning is retained.
 either first or second order. Although, as is 
discussed in the Introduction, there are works showing that even classical 
phase transitions could be of zeroth order. Even more so is allowed for 
such a non-classical transition as a nucleation transition.

\vskip 3cm

\vskip 2cm

%Figure 2

\vskip 2cm

%Figure 3

\vskip 2cm

%Figure 4

\vskip 2cm

%Figure 5

\vskip 2cm

%Figure 6

\vskip 2cm

%Figure 7

\vskip 2cm

%Figure 8

\end{document}